\documentclass[prc,aps,12pt,final,notitlepage,oneside,onecolumn,
nobibnotes,nofootinbib,superscriptaddress,noshowpacs]{revtex4-1}
\usepackage{graphicx,colordvi}
\usepackage{amssymb,amsmath}
\usepackage{mathtext}
\usepackage{bm}
\usepackage{array}
\usepackage{calc}
\usepackage{ifthen}
\usepackage{epsfig}

\def\be{\begin{equation}}
\def\ee{\end{equation}}
\def\bea{\begin{eqnarray}}
\def\eea{\end{eqnarray}}

\def\siml{\;\hbox{\kern.1em \lower.7ex \hbox{$\sim$} \kern-1.12em
 \raise.5ex \hbox{$<$} \kern.1em}}
\def\simg{\;\hbox{\kern.1em \lower.7ex \hbox{$\sim$} \kern-1.12em
 \raise.5ex \hbox{$>$} \kern.1em}}

\begin{document}

\title{ CHAOTICITY AND SHELL EFFECTS IN THE \\ 
NEAREST-NEIGHBOR DISTRIBUTIONS}

\author{J.P. Blocki} 
\affiliation{\it National Center for Nuclear Research,
Otwock 05-400, Poland}
\author{\footnotesize A.G. Magner\footnote{magner@kinr.kiev.ua}}
\affiliation{\it Institute for Nuclear Research,  Kyiv 03680, Ukraine}

\begin{abstract}

Statistics of the single-particle levels in a deformed
Woods-Saxon potential is analyzed in terms of the Poisson and Wigner
nearest-neighbor distributions for several deformations 
and multipolarities of its surface distortions. 
We found the significant
differences of all the distributions with a fixed value of the angular momentum
projection of the particle, more closely
to the Wigner distribution, in contrast to the full spectra with Poisson-like 
behavior. Important shell effects are observed in 
the nearest neighbor spacing distributions, 
the larger the smaller deformations of the surface 
multipolarities.

\end{abstract}

\bigskip

\centerline{\today}


\maketitle

\section{INTRODUCTION}

The microscopic many-body interaction of particles of the Fermi systems such 
as heavy nuclei 
is rather complicated. Therefore, several 
theoretical approaches 
to the description of the Hamiltonian which are based on the 
statistical properties of its 
discrete levels are applied for solutions of the realistic problems. 
For a quantitative measure of the
degree of chaoticity of the many-body forces,  
the statistical distributions
of the spacing between the nearest neighboring levels
were introduced, first of all in relation to the so called Random Matrix
Theory \cite{porter,wigner,brody,mehta,aberg}. 
Integrability (order) of the system was associated usually to the 
Poisson-like exponentially decreasing dependence on the spacing
variable with a maximum at zero 
while chaoticity was connected more to the Wigner-like behavior 
with the 
zero spacing probability at zero but with a maximum at some finite value of 
this variable.

On the other hand, many dynamical problems, in particular, 
in nuclear physics can be reduced 
to the collective motion of independent particles in a mean field with a 
relatively sharp time-dependent edge called usually as the effective 
surface within the microscopic-macroscopic 
approximation \cite{myersswiat,strut}. 
We may begin with the basic ideas of Swiatecki and his 
collaborators
\cite{myersswiat,wall,blsksw1,blshsw,blsksw2,jasw,maskbl}.
In recent years it became apparent that the collective nuclear
dynamics is very much related to the nature of the nucleonic motion.
This behavior of the nucleonic dynamics is important in physical processes
like fission or heavy ion collisions where a great amount of the
collective energy is dissipated into a chaotic nucleonic motion.
We have to mention here also very intensive studies of 
the one-body dissipative phenomena described largely through the 
macroscopic wall formula (w.f.) for the excitation 
energy \cite{wall,blsksw1,blsksw2,blshsw,jasw,maskbl} and also 
quantum results \cite{blsksw1,blsksw2,maskbl,kiev2010}. 
The analytical w.f. was
suggested originally
in Ref.\ \cite{wall} on the basis of the Thomas-Fermi approach.
It was re-derived in many works
based on semiclassical and quantum arguments, 
see Refs\ \cite{koonrand,maggzhfed,gmf} for instance. 
However, some 
problems in the analytical study of a multipolarity dependence
of the smooth one-body friction and its oscillating corrections as functions
of the particle number should be still clarified. 
In particular, we would like to emphasize the 
importance of the transparent classical picture 
through the Poincare 
sections and Lyapunov exponents showing the order-chaos 
transitions \cite{arvieu,blshsw,heiss,heiss2} and also 
quantum results for the excitation 
energy \cite{blsksw2,maskbl,kiev2010}.  
Then, the peculiarities of the excitation energies for 
many periods of the oscillations of the classical
dynamics were  discussed for several Legendre 
polynomials, see Ref.\ \cite{kiev2010}.
as the classical 
measures of chaoticity. The shell correction
method \cite{strut,FH} 
 was successfully used to describe the shell effects
in the nuclear deformation
energies as functions of the particle numbers.
This is important also 
for understanding analytically the origin of the isomers in fission 
within the periodic
orbit theory (POT) \cite{myab}. We should expect also that 
the deviations of the level density near the Fermi surface, 
like shell effects,
from an averaged constant should influence essentially the nearest
neighbor spacing distribution (NNSD).
For a further study
of the order-chaos properties of the Fermi systems, it might be worth 
to apply 
the statistical methods of the description of the single-particle (s.p.) levels 
of a mean-field Hamiltonian within microscopic-macroscopic 
approaches (see for instance Refs\ \cite{heiss1,heiss3,nazm1,nazm2}).

The statistics of the 
spacing between the neighboring levels and their relation to the shell effects
depending on the specific properties of the s.p.
spectra, as well as the  multipolarity and deformation of the shape surfaces 
should be expected. The quantitative measure of the order
(or symmetry) can be the number of the single-valued integrals of motion, 
except for
the energy (degree of the degeneracy of the system, see also 
Refs\ \cite{strutmag,smod,myab,brbhad}). If the energy is the only one
single-valued integral of motion one has the completely chaotic system
\cite{gutz}. For the case of any such additional integral of motion, say
the angular momentum projection for the 
azimuthal symmetry, one finds the symmetry enhancement that is important
for calculations of the level density as the basic s.p. 
characteristics.  

Our purpose now is to look at the  
order-chaos properties of the s.p. levels in terms of the
Poisson and Wigner distributions with focus to their
dependence on the multipolarities, equilibrium deformations 
and shell effects in relation to
the integrability of the Hamiltonian through the comparison between the 
spectra with the fixed angular momenta of particles and full
for the Woods-Saxon potential.

\section{SPECTRA AND LEVEL DENSITIES}

We are going now to study the statistical properties of the 
s.p. spectra of the eigenvalue problem,
\begin{equation}
H \phi_i = \varepsilon_i \phi_i ,\qquad\qquad H=T+V,
\label{eigenprob}
\end{equation}
where $H$ is a static mean-field Hamiltonian 
with the operator of the kinetic
energy $T$ and deformed axially-symmetric Woods-Saxon (WS) potential, 
\begin{equation}
V\equiv V_{WS}\left({\bf r}\right)=-
\frac{V_0}{1+ \exp\left\{\left[r-R(\theta)\right]/a\right\}}\;,
\label{wspot}
\end{equation}
$r,\theta,\varphi$ are the spherical coordinates of the vector ${\bf r}$.
Following Refs\ \cite{blsksw1,blshsw,jasw,maskbl},
the 
shape of the WS-potential surface is defined by the effective 
radius $R(\theta)$ given by:
\begin{equation}
R(\theta)=\frac{R^{}_0}{\lambda}\left[1+\alpha\; \sqrt{\frac{4\pi}{5}}\; 
Y^{}_{n0}\left(\theta\right)
+\alpha_1 \sqrt{\frac{4\pi}{3}}\; 
Y^{}_{10}\left(\theta\right)\right].
\label{radst}
\end{equation}
Here, $\lambda$ is a normalization factor ensuring 
volume conservation, and $\alpha^{}_1$ stands 
 for keeping a position of the center of mass 
for odd multipolarities and
$R^{}_0$ is the radius of the
 equivalent sphere, $Y^{}_{n0}(\theta)=
\sqrt{(2n+1)/4\pi}\;P^{}_n\left(\cos\theta\right)$ are the spherical functions. 
$P^{}_n\left(\cos\theta\right)$ are the Legendre polynomials and
$\alpha$ is the deformation parameter
independent on time.
For diagonalization of the Hamiltonian with the WS potential (\ref{wspot}), 
the expansion
over a basis of the deformed harmonic oscillator is used as shown in 
Ref.\ \cite{maskbl}.

\begin{figure}
\begin{center}
\includegraphics[width=0.45\textwidth,clip]{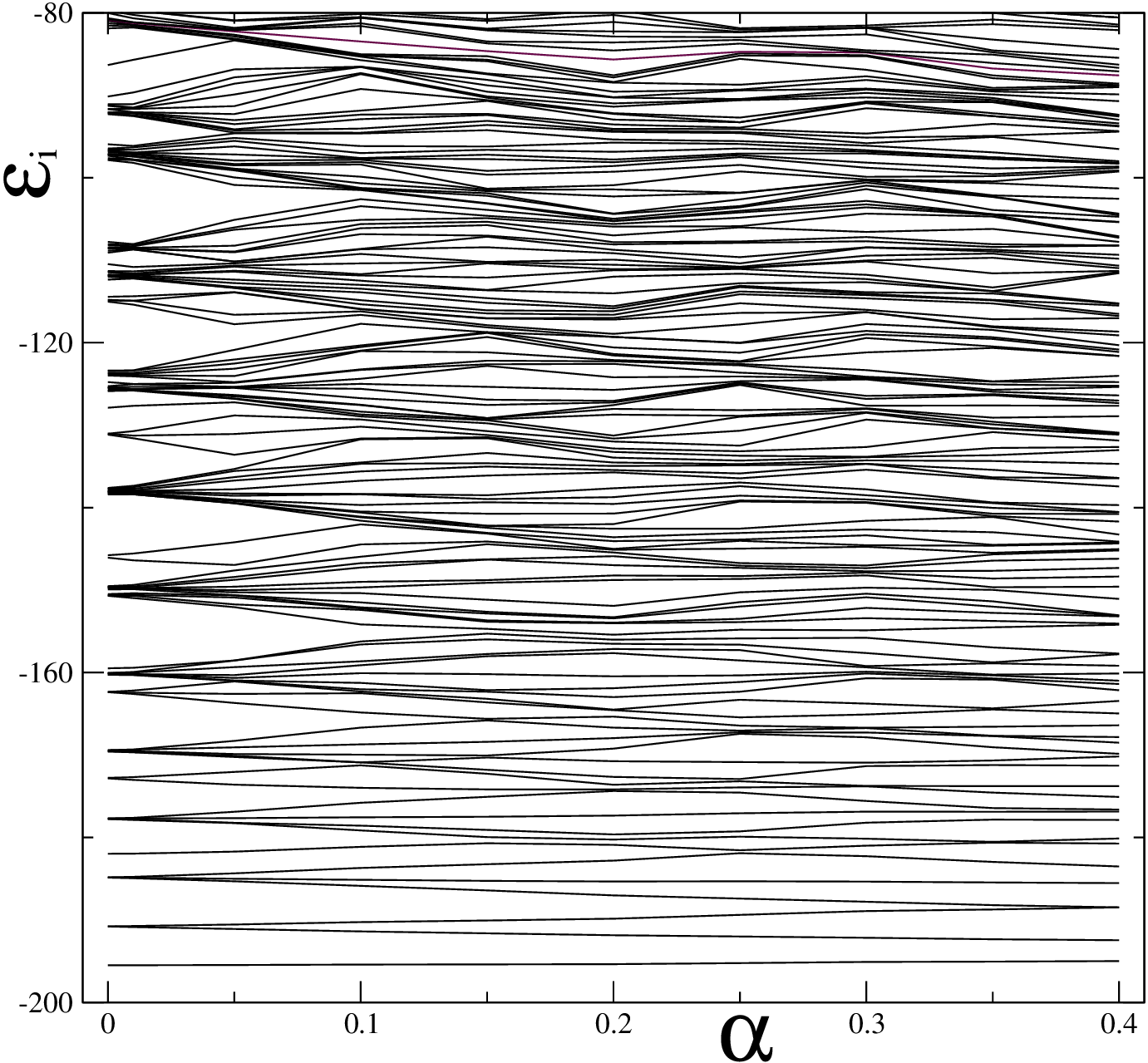}
\hspace{0.1cm}
\includegraphics[width=0.45\textwidth,clip]{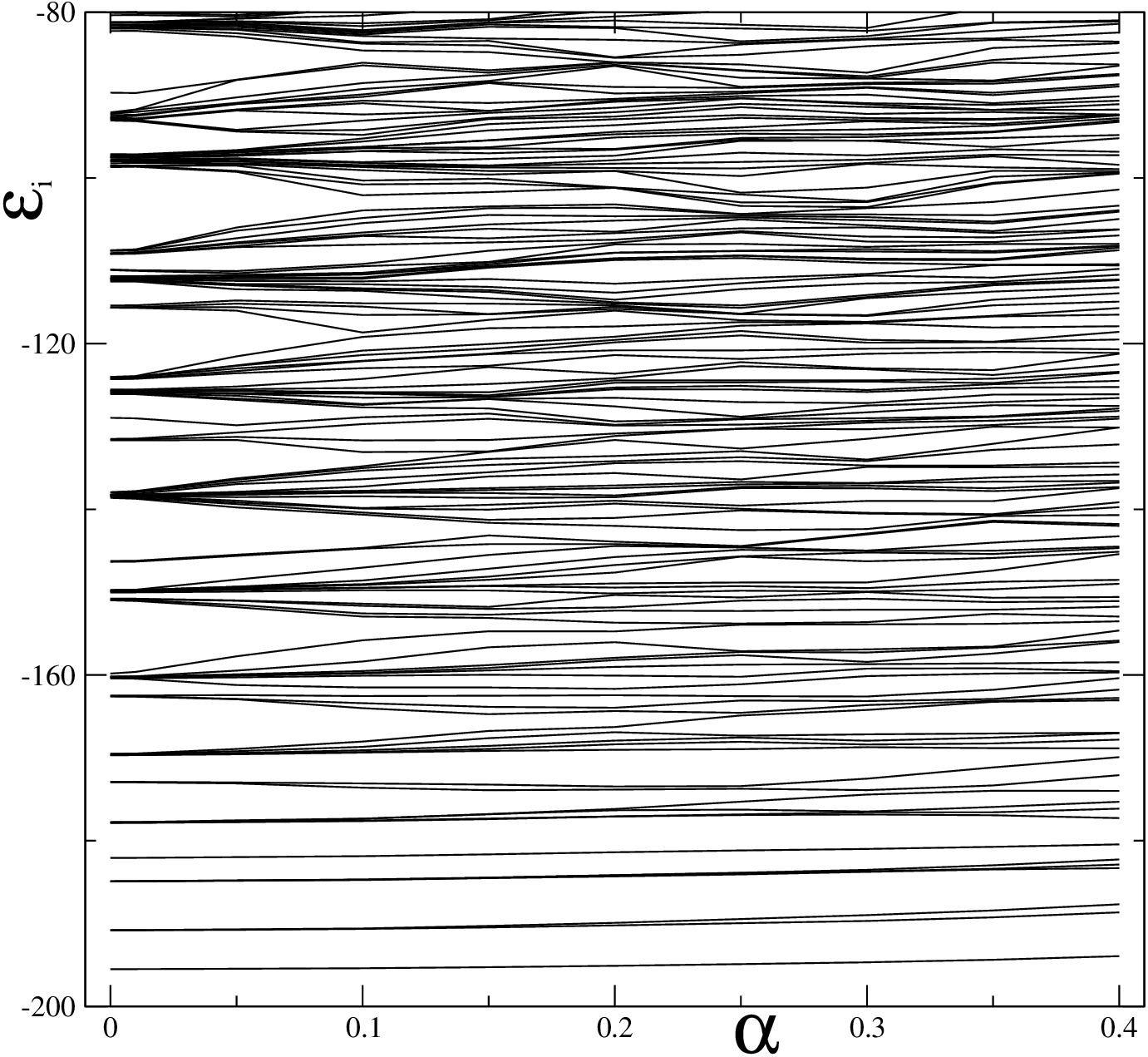}
\end{center}
\caption{ 
The s.p. energy
levels $\varepsilon_i$ 
in the WS potential ($V_0=200 $ MeV, $R_0=6.622$ fm, $a=0.1$ fm)
as function of the deformation $\alpha$ for the $P_2$ (left) and 
$P_5$ (right) shapes (Eq.\ (\ref{radst})).}
\label{fig1}
\end{figure}

\begin{figure}
\begin{center}
\includegraphics[width=0.45\textwidth,clip]{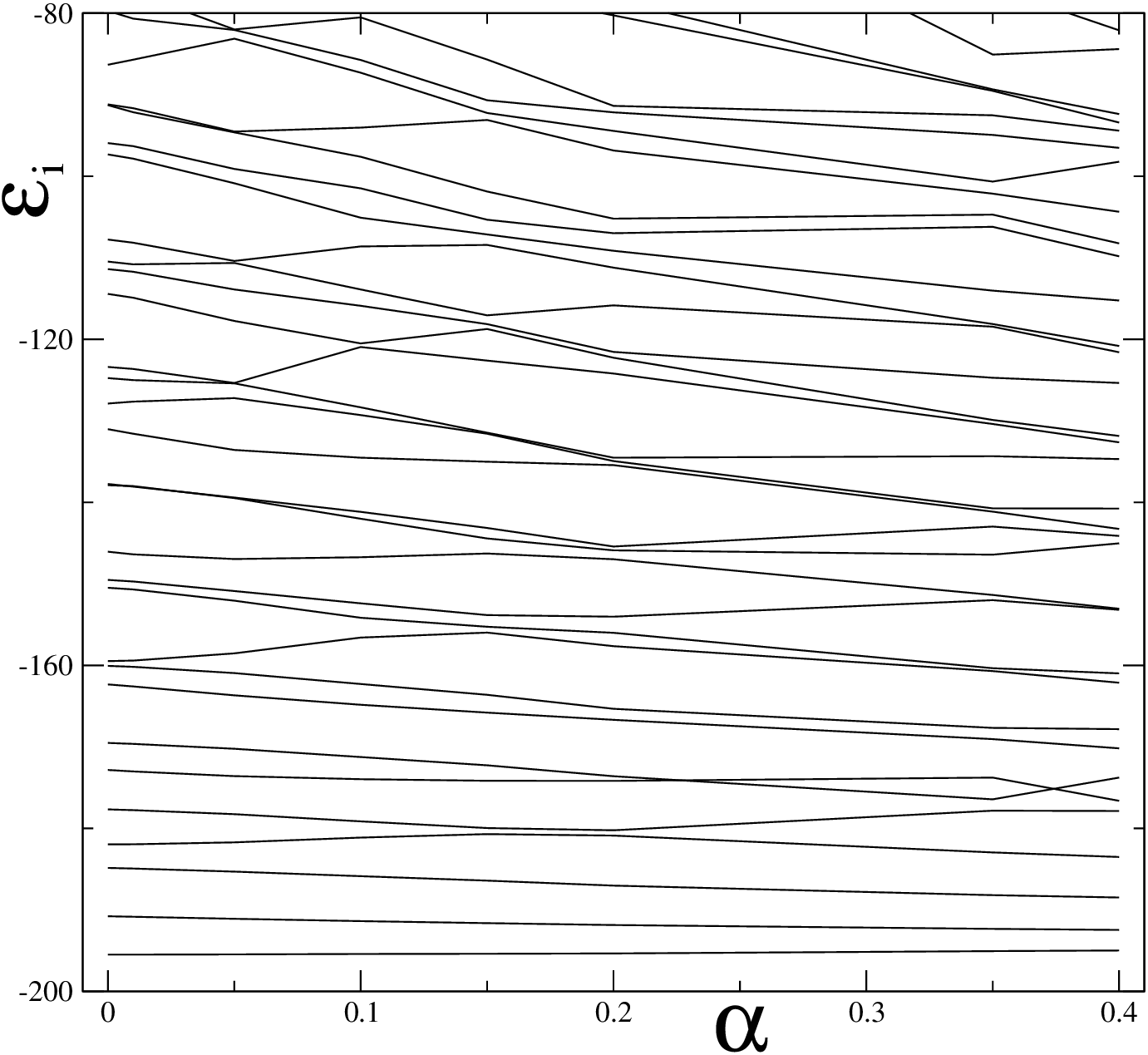}
\hspace{0.1cm}
\includegraphics[width=0.45\textwidth,clip]{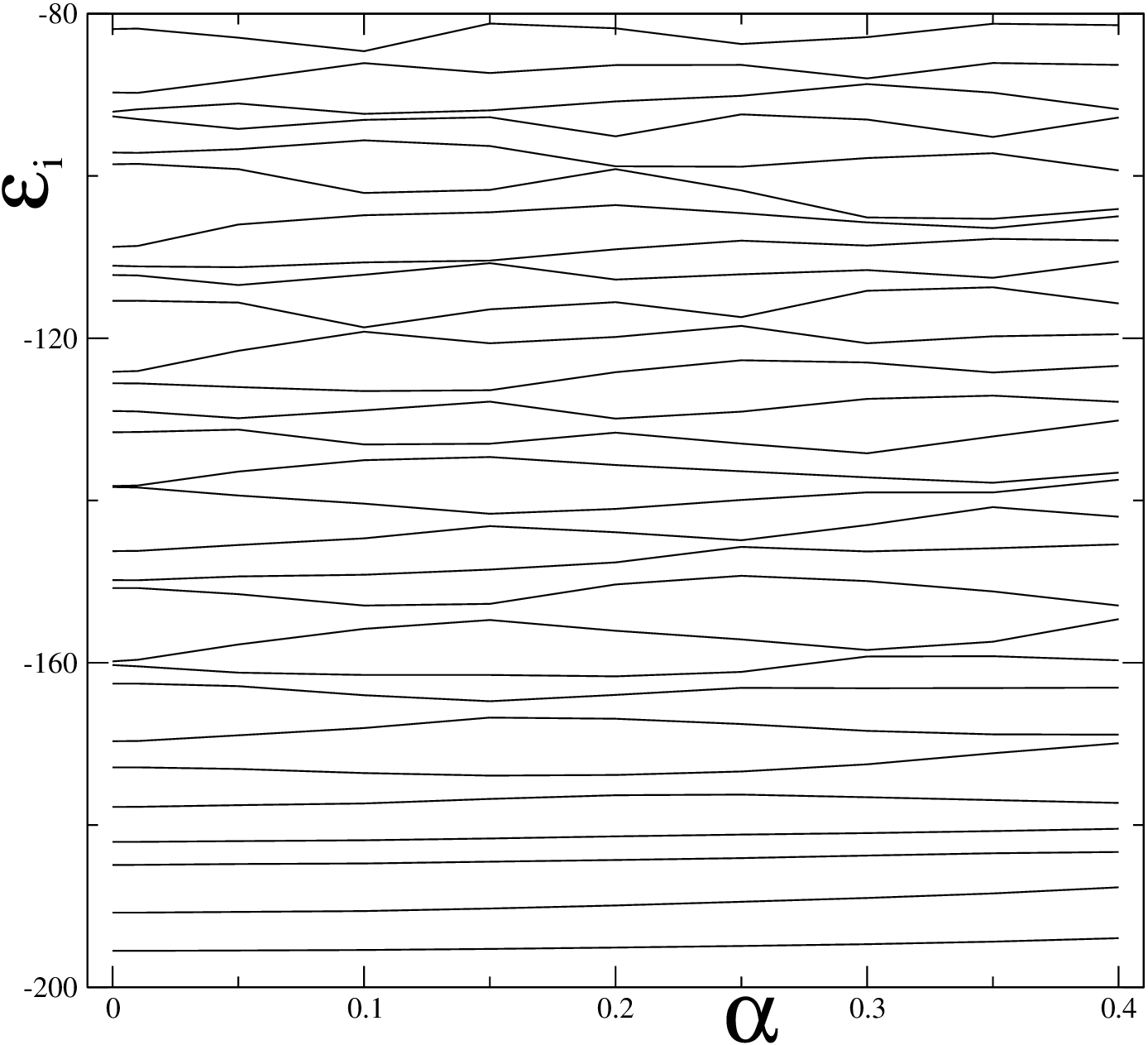}
\end{center}
\caption{ 
The same s.p. spectrum of levels as in Fig.\ \ref{fig1} but 
with the projection of the angular momentum $m=0$.} 
\label{fig2}
\end{figure}

Figs\ \ref{fig1} and \ref{fig2} show two examples for the full spectra of the 
s.p. energies 
$\varepsilon_i$ and for the fixed angular momentum 
projection $m=0$ 
versus the deformation  parameter $\alpha$ for the $P_2$ and $P_5$ shapes,
respectively.
The spectra for $P_3$ and $P_4$ are very similar to the $P_5$ 
case and therefore,
they are not shown. As seen from Fig.\ \ref{fig1},
there are clear shell effects in the full spectra at small deformations, 
approximately at $\alpha \siml 0.1$ for all multipolarities 
from the $P_2$ shape to 
the $P_5$ one. With increasing deformation $\alpha$, the shell gaps 
become less pronounced and slowly changed in the region  
$\alpha \approx 0.1-0.4$ for all these multipolarities. Much more differences 
can be found in comparison of Fig.\ \ref{fig1} for  full spectra and 
Fig.\ \ref{fig2} for $m=0$ levels only. The shell effects are seen here too but
much less pronounced. 
The spectrum of levels with $m=0$ 
becomes more uniform with increasing multipolarity $n$.

\begin{figure}
\begin{center}
\includegraphics[width=0.45\textwidth,clip]{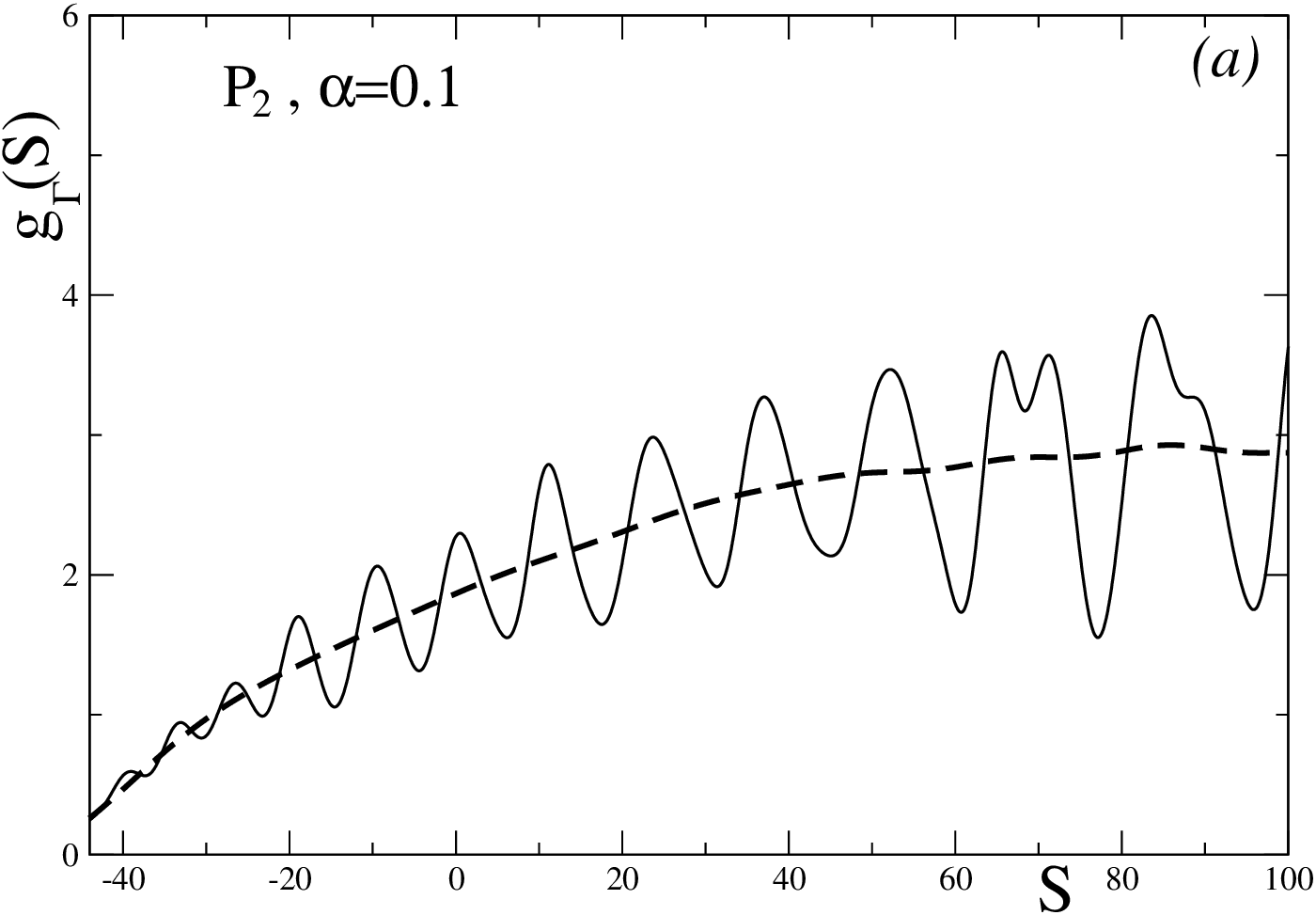}
\hspace{0.1cm}
\includegraphics[width=0.45\textwidth,clip]{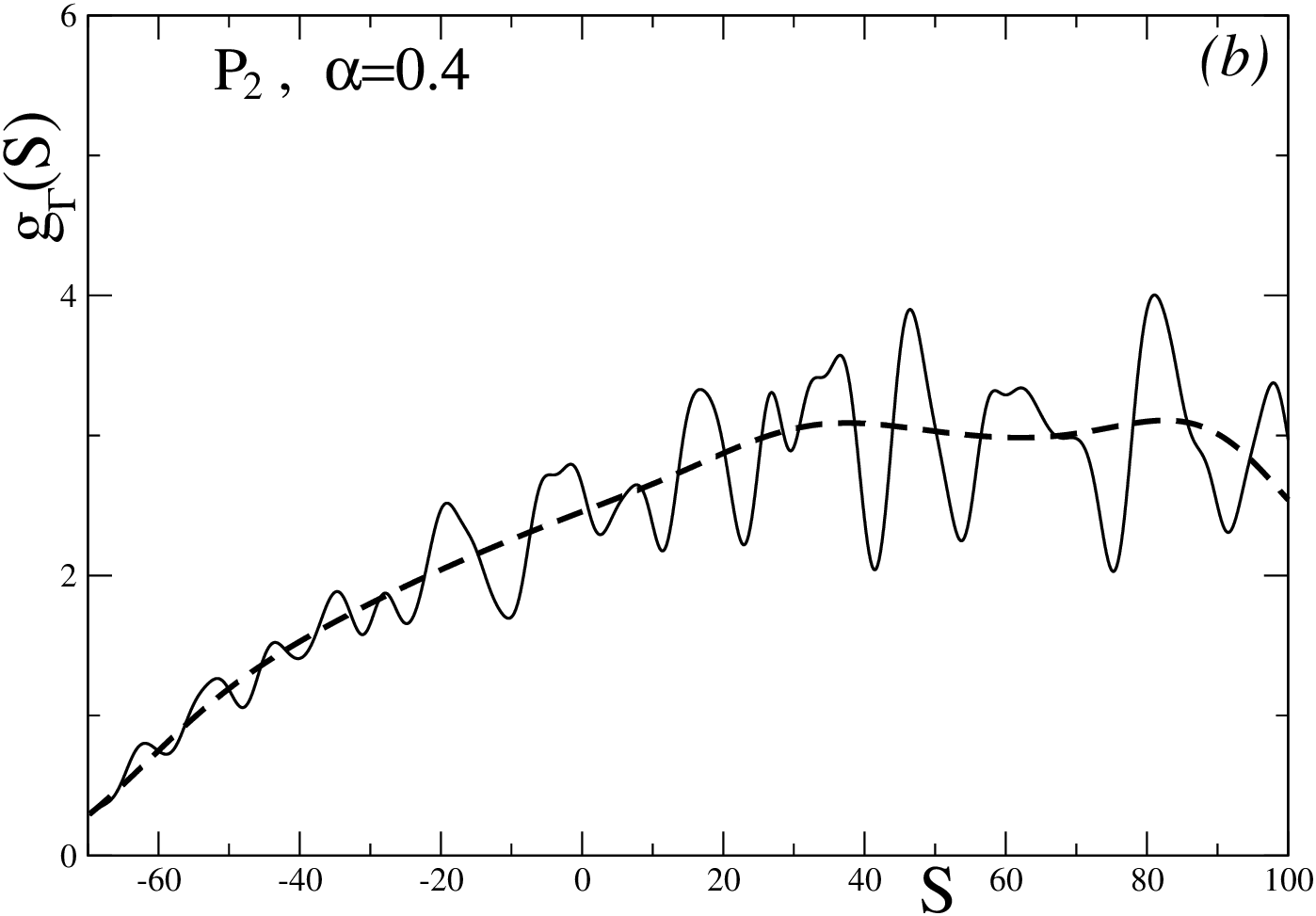}

\includegraphics[width=0.45\textwidth,clip]{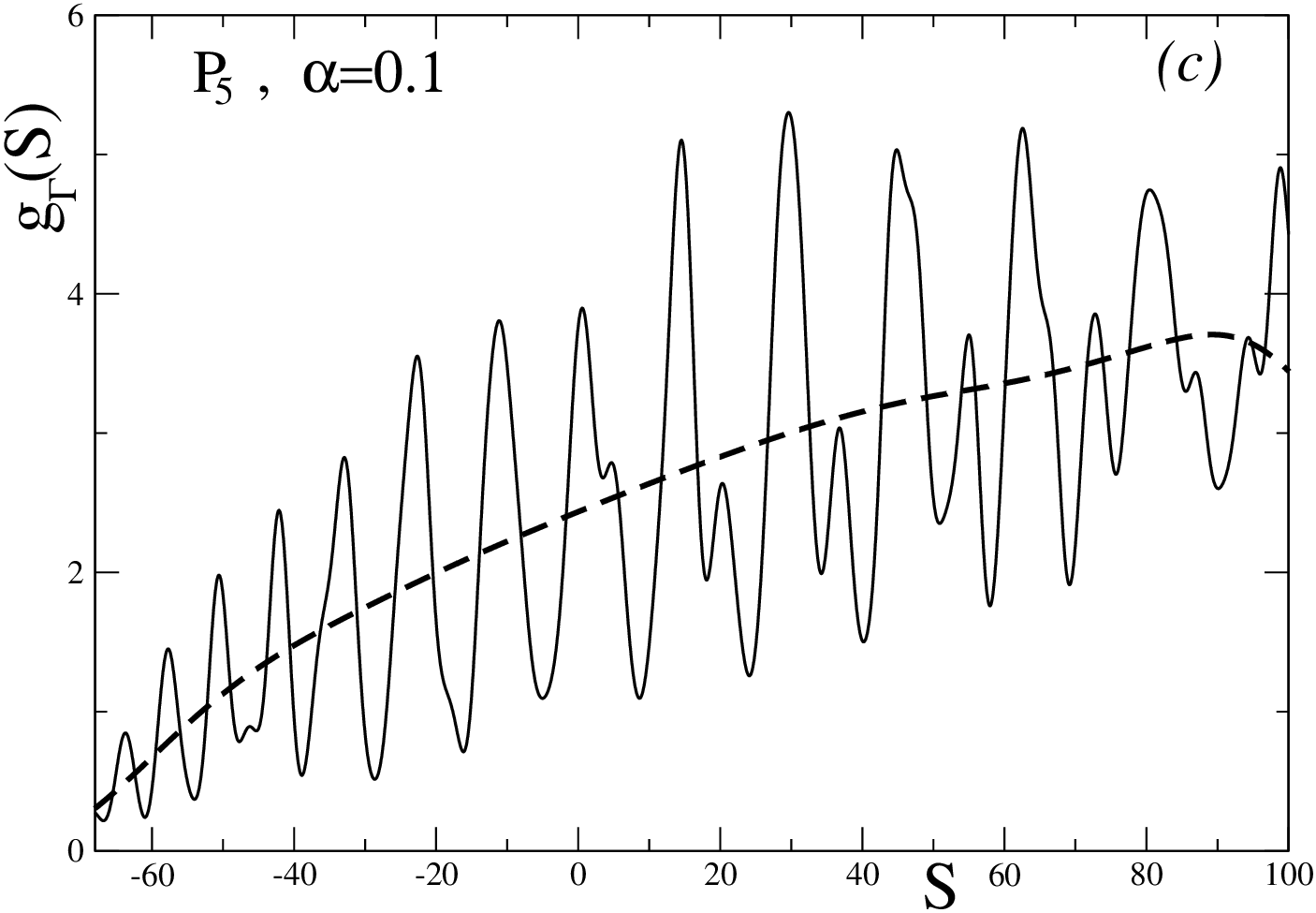}
\hspace{0.1cm}
\includegraphics[width=0.45\textwidth,clip]{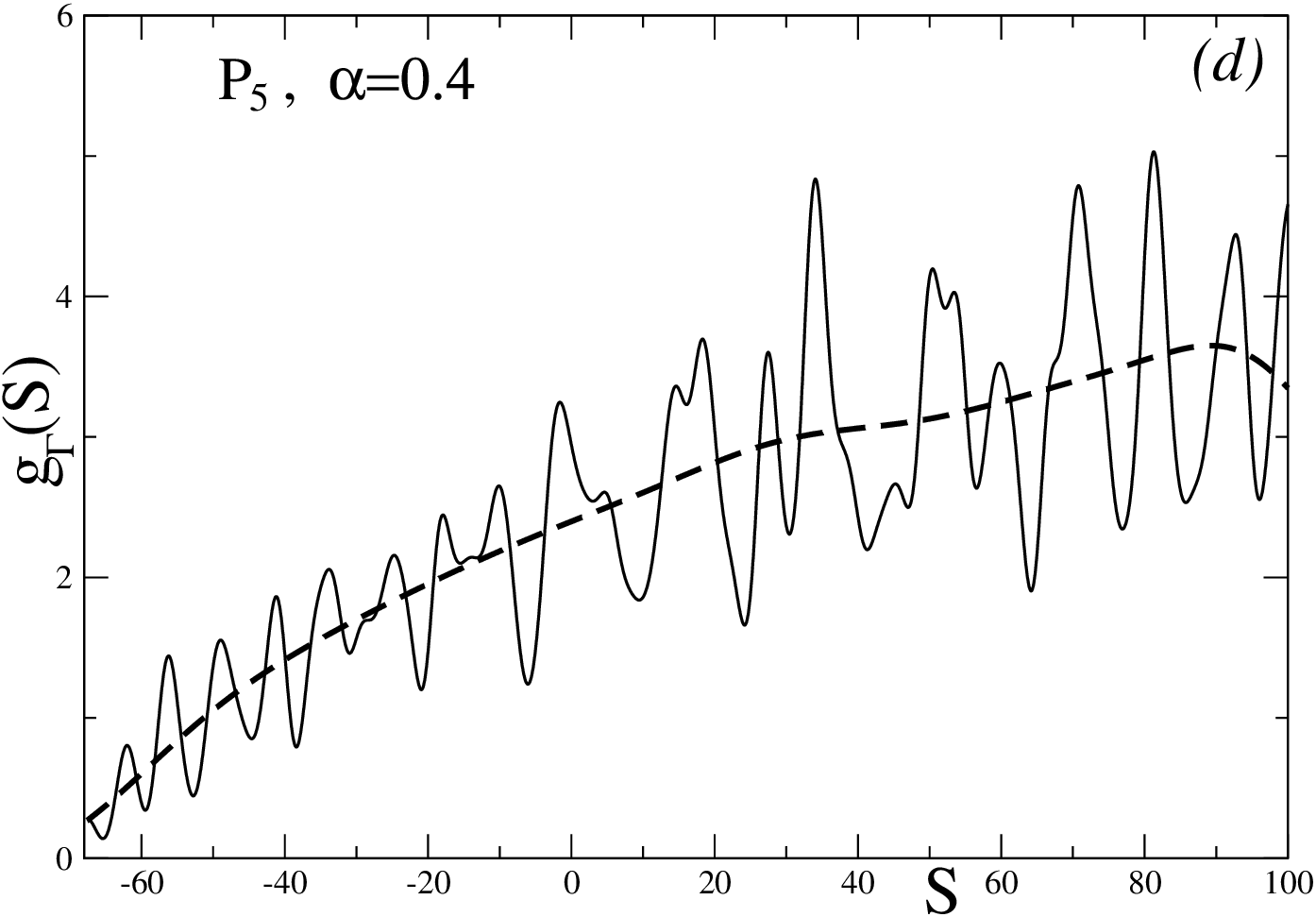}
\end{center}
\caption{ 
The level densities $g_\Gamma(S)$ (Eq.\ (\ref{totlevden})) 
as a function 
of the energy
counted from the Fermi level ($S=0$) for a given particle number $N$
for spectra of the s.p. levels of Fig.\ \ref{fig1} 
for the $P_2$ $(a,b)$  and $P_5$ 
$(c,d)$ shapes at the
small $\alpha=0.1$ (left, $a,c$) and larger   $\alpha=0.4$ (right,$b,d$) 
deformations;
dashed is the smooth density $\tilde{g}(S)$ (Eq.\ (\ref{totlevden})); 
solid is the total density $g_\Gamma(S)$ (Eq.\ (\ref{totlevden}))
($\Gamma=3$ MeV, $M=0$ for 
$\delta g_\Gamma(S)$).}
\label{fig3}
\end{figure}

\begin{figure}
\begin{center}
\includegraphics[width=0.45\textwidth,clip]{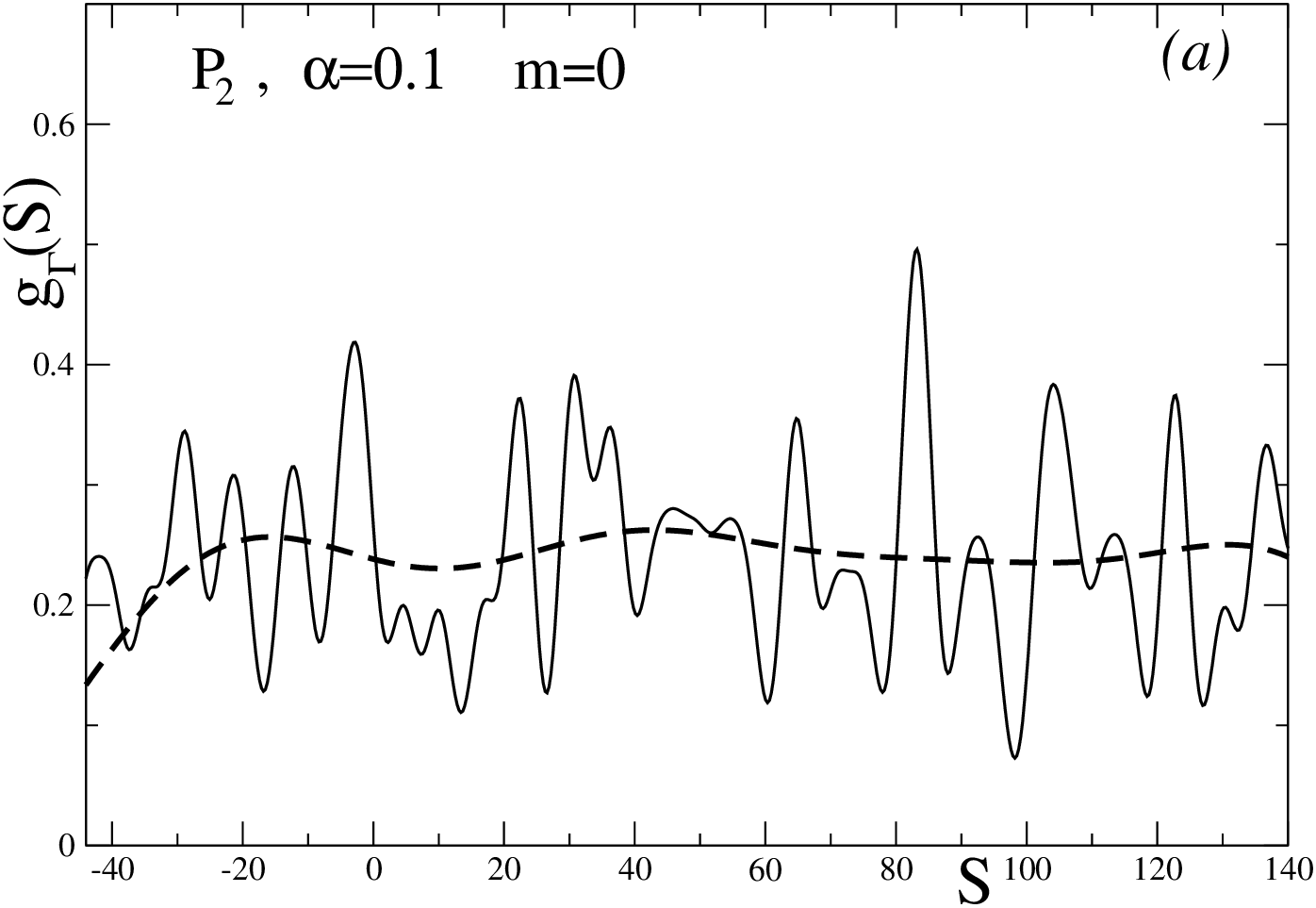}
\hspace{0.1cm}
\includegraphics[width=0.45\textwidth,clip]{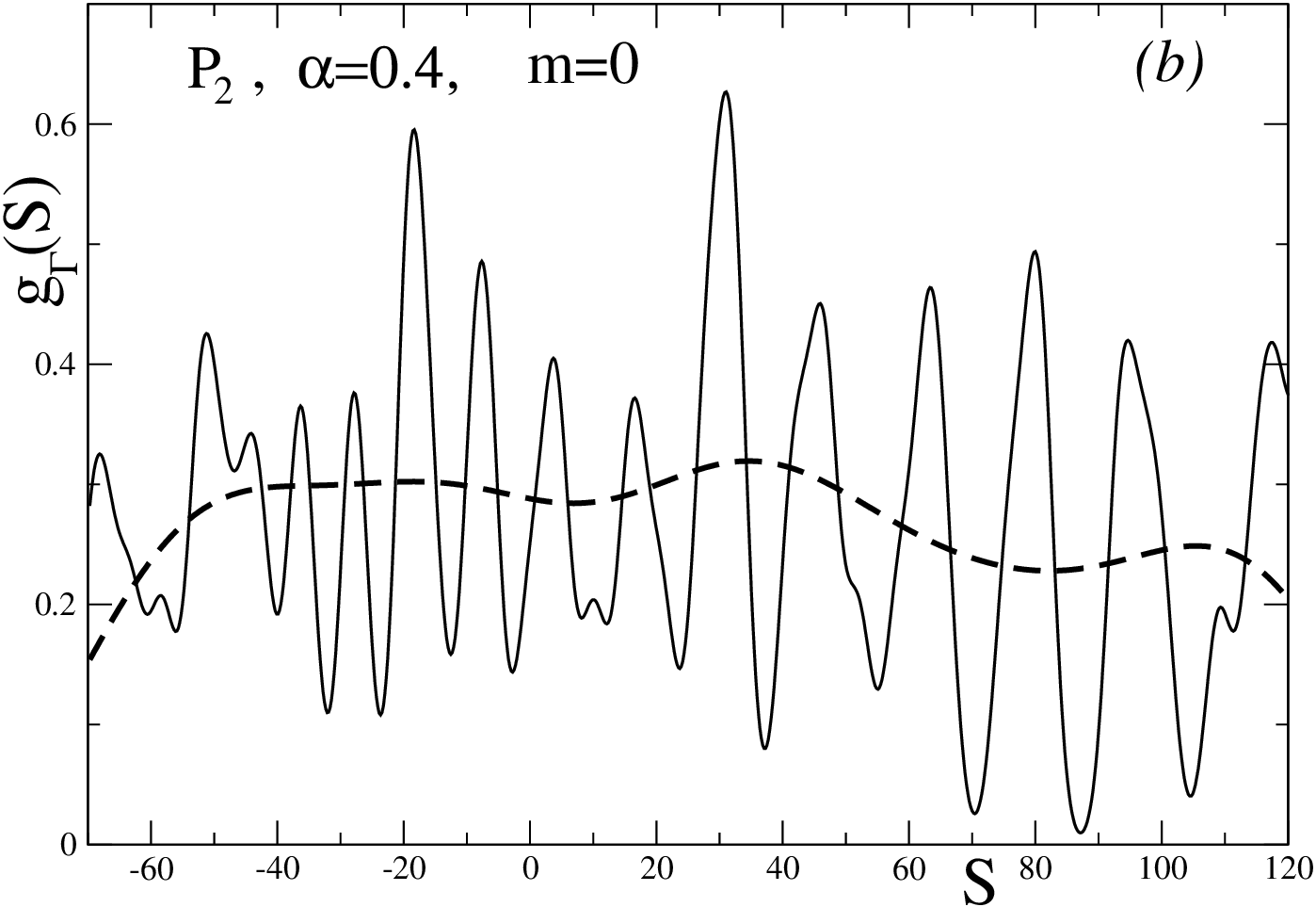} 

\includegraphics[width=0.45\textwidth,clip]{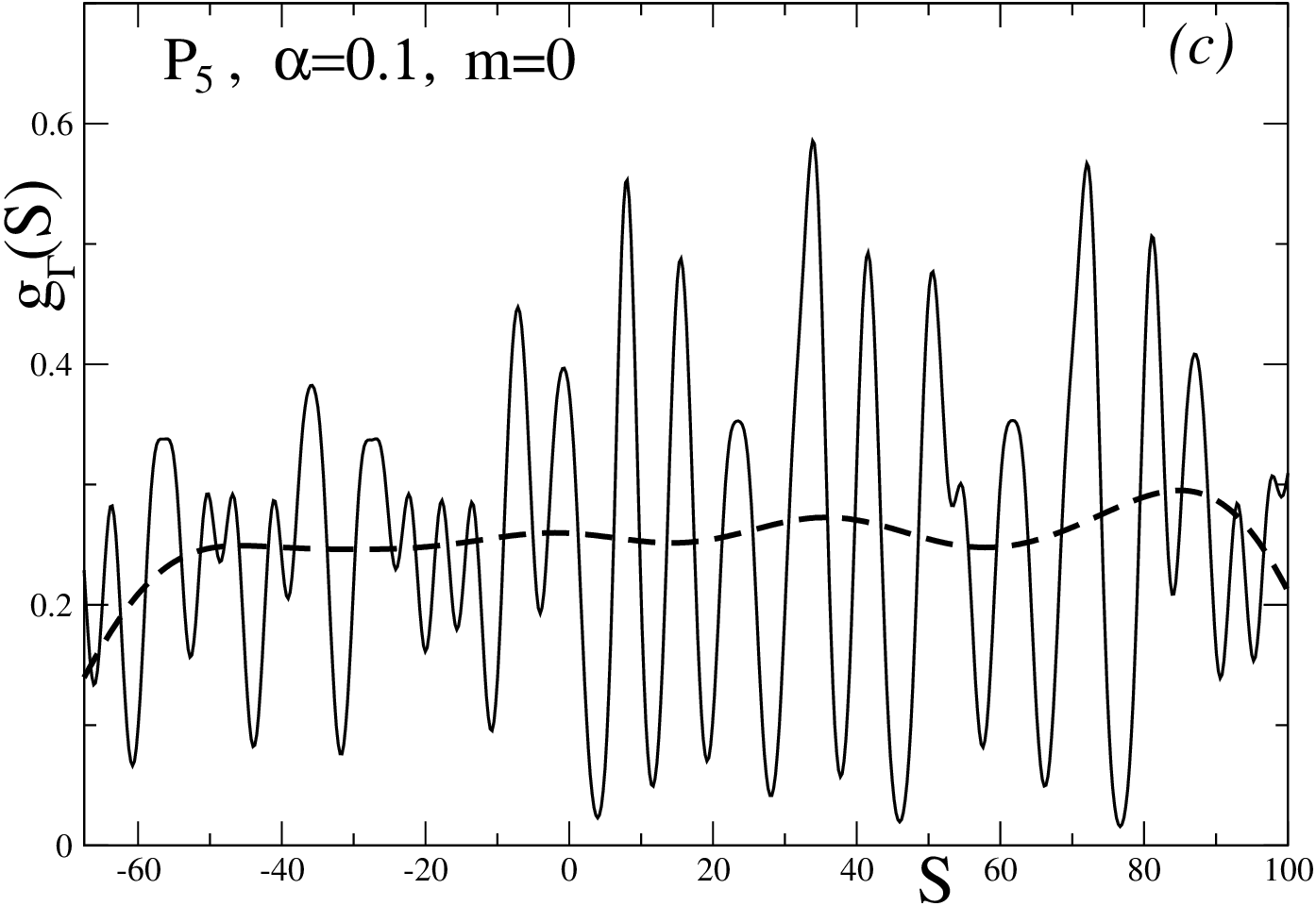}
\hspace{0.1cm}
\includegraphics[width=0.45\textwidth,clip]{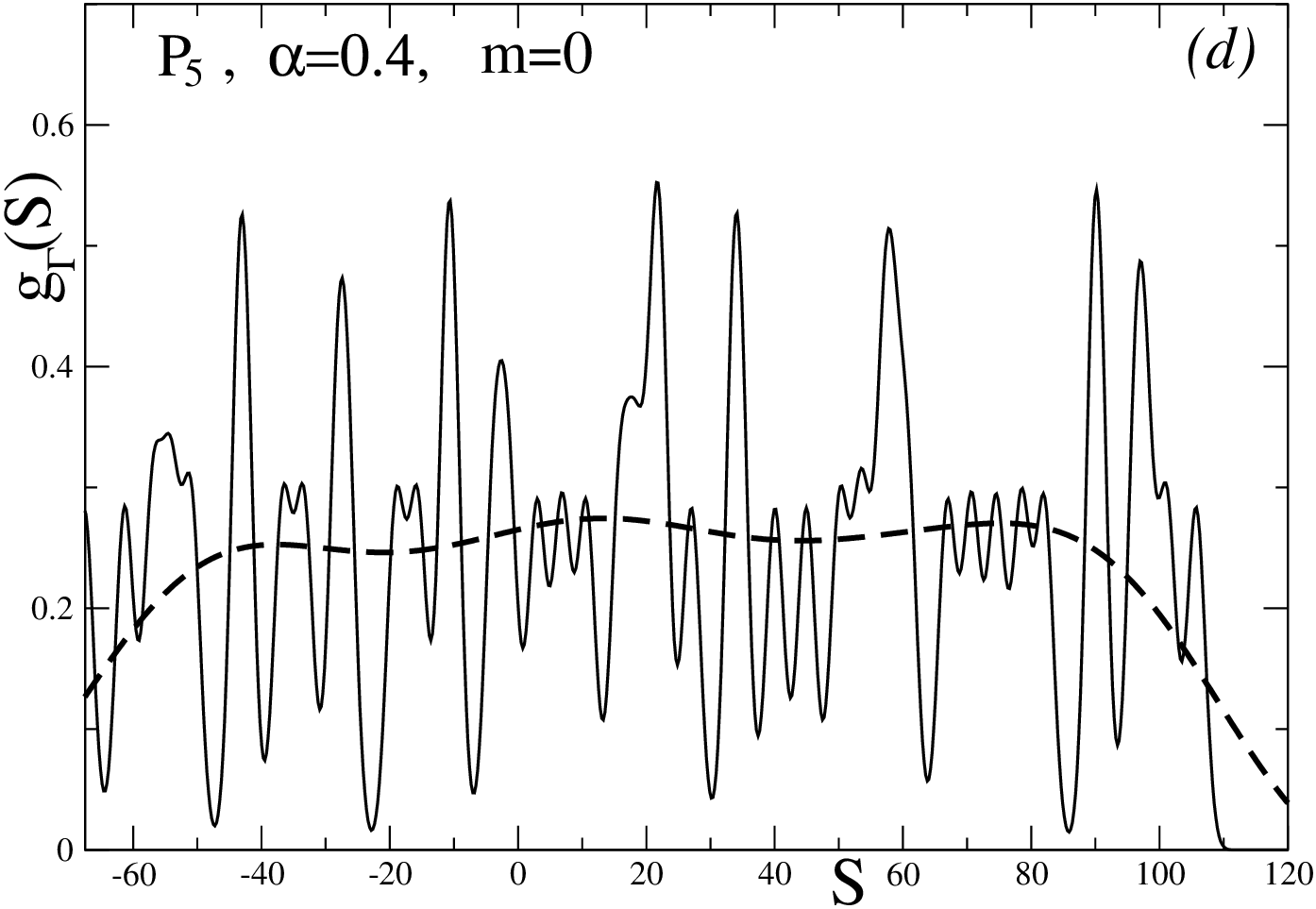}
\end{center}
\caption{
 The same as in Fig.\ \ref{fig3} but for levels with the fixed projection 
of the angular momentum  $m=0$.}
\label{fig4}
\end{figure}

The key quantity for calculations of the NNSD
$P(S)$ is the level density $g(\varepsilon)$, see Appendix A and Refs\ 
\cite{wigner,porter,brody,aberg,mehta}.
For these calculations one may apply the Strutinsky shell correction
method writing
\begin{equation}
g^{}_{\Gamma}(\varepsilon)=\tilde{g}(\varepsilon) + 
\delta g^{}_\Gamma(\varepsilon).
\label{totlevden}
\end{equation}
The smooth part $\tilde{g}(\varepsilon)$ is defined by the Strutinsky
smoothing procedure 
\cite{strut,FH}.
The so called plateau condition (stability
of values of the smooth level density $\tilde{g}$ as function of the 
averaging parameters: 
Gaussian width $\Gamma$ and the degree of the correction polynomial
$M$ takes place at $\Gamma=20-40$ MeV and $M=4-8$). 
Figs\ \ref{fig3} and \ref{fig4} for full spectra
and for fixed angular momentum projection $m=0$ show the 
typical examples of the level densities
(smooth component and the total density with the oscillating part) for the 
same degree of the Legendre polynomials $n=2$ and $5$ at  
deformations $\alpha=0.1$ and $0.4$,
in correspondence with spectra presented in Figs\ \ref{fig1} and \ref{fig2}, 
respectively.
In Fig.\ \ref{fig4} for the case of the specific $m=0$ levels, one has somewhat
larger Gaussian width parameters of the smooth level density $\tilde{g}$ in the
total density (Eq.\ (\ref{totlevden})) than
those for the full spectra in Fig.\ \ref{fig3}. 
As seen clearly from Figs\ \ref{fig3} and \ref{fig4} 
the smooth
level density for the $m=0$ spectra is
more flat (besides of relatively small remaining oscillations
because of much less levels with the fixed $m=0$), 
as compared to the full-spectra results. This more flat behavior for
the fixed $m$ value is due to the loss of the symmetry.
Therefore one expects the system with the fixed angular momentum $m$ to
be more chaotic. The differences between Fig.\ \ref{fig3} and 
Fig.\ \ref{fig4} are quite
remarkable. On the other hand differences between pictures within 
Fig.\ \ref{fig3} or
Fig.\ \ref{fig4} are 
less notable and as one can see shell effects are still remaining 
at bigger deformations and multipolarities. 

\section{NEAREST-NEIGHBOR SPACING DISTRIBUTIONS}

Following the review paper \cite{brody}, the  
distribution ${\cal P}(S)$ for the probability of finding the 
spacing $S$ 
between the nearest neighboring levels 
is given by (see also Refs\ \cite{wigner,porter,aberg,mehta} and Appendix A)
\begin{equation}
{\cal P}(S)=g(S)\;
\exp\left(-\int_0^S g(x)\;\mbox{d}x\right)/\aleph.
\label{genwig}
\end{equation}
The key quantity $g(S)$ can be considered as the density of the s.p. levels 
counted from a given energy, say, 
the Fermi energy $E_F$. $D$ is a mean uniform 
distance between neighboring levels 
so that $1/D$ is the mean density of levels.
$\aleph$ is the normalization factor for large enough maximal value
of $S$, $S_{max}$,
\begin{equation}
\aleph=\int_0^{S_{max}}\mbox{d}x g(x) 
\;exp\left(-\int_0^x g(y)\mbox{d}y\right)/D.
\label{norm}
\end{equation}
This normalization factor $\aleph$ can be found from 
the normalization conditions:
\begin{equation}
\int \mbox{d}x {\cal P}(x)=
\int \mbox{d}x x {\cal P}(x)= 1\;.
\label{normcond}
\end{equation}
[Notice that for convenience
we introduced the dimensionless probability ${\cal P}$ in contrast to
that of Ref.\ \cite{brody} denoted as $P(S)$, see Eq.\ (1.3) there.]

The Poisson law follows if we take constant for the level density, $g(S)=1/D$,
in Eq.\ (\ref{genwig}), 
\begin{equation}
{\cal P}(S)
=\exp\left(-S/D\right)\;.
\label{pois}
\end{equation}
Wigner's law follows from the assumption of the linear level density, 
proportional to $S$,
\begin{equation}
{\cal P}(S)
=\left(\pi S/2 D\right)\;\exp\left(-\pi S^2/4 D^2\right)\;.
\label{wig}
\end{equation}
Both distributions are normalized to one for large enough maximal value of $S$,
$S_{max}=\infty$ to satisfy Eq.\ (\ref{normcond}).

The level density in fact is not a constant or $\propto S$.
The combination of the Poisson and Wigner distributions
was suggested in Ref.\ \cite{berrob} by introducing one parameter.
For our purpose to keep a link with the properties of the
level density, like smooth and shell components \cite{strut}, it is 
convenient to define the probability
${\cal P}(S)$ (Eq.\ (\ref{genwig})) for a general linear level-density function
through two parameters ${\cal A}$ and ${\cal B}$,
\begin{equation}
g(S)=\left({\cal A} + {\cal B}S/D\right)/D\;.
\label{denlin}
\end{equation}
Substituting Eq.\ (\ref{denlin}) into the general formula
 (Eq.\ (\ref{genwig})) one obtains explicitly the analytical result in terms of
the standard error functions, 
$\mbox{erf}(z)=2\int_0^z \mbox{d}x\;\exp(-x^2_{})/\sqrt{\pi}$,
\begin{equation}
{\cal P}(S)=\left(1+{\cal B} \xi/{\cal A}\right)\;
\exp\left(-{\cal B} \xi^2_{}/2 - {\cal A} \xi\right)/
\left[\aleph_0 + {\cal B}\;\aleph_1/{\cal A}\right]\,, 
\label{pslinPW}
\end{equation}
\begin{eqnarray}
\aleph_0&=&
\int_0^{\cal C} \mbox{d}\xi \;
\exp\left(-\frac{{\cal B}}{2} \xi^2 - {\cal A}\xi\right)=
\sqrt{\frac{\pi}{2{\cal B}}}\;\exp\left(\frac{{\cal A}^2}{2{\cal B}^2}\right)\;
\mbox{erf}\left(\frac{{\cal A}+{\cal B}{\cal C}}{\sqrt{2{\cal B}}}\right),
\nonumber\\
\aleph_1&=&
\int_0^{\cal C} \mbox{d}\xi \xi\;\exp\left(-\frac{{\cal B}}{2} \xi^2 - 
{\cal A} \xi\right)=
-\frac{1}{{\cal B}}\left[\exp\left(-\frac{{\cal B}}{2} {\cal C}^2 - 
{\cal A} {\cal C}\right)+ {\cal A}\;\aleph_0\right].
\label{pslin}
\end{eqnarray}
where  $\xi=S/D$, ${\cal C}=S_{max}/D$ is the maximal value of $\xi$.
For large ${\cal C} \rightarrow \infty$ one has simply 
$\aleph_0 \rightarrow \sqrt{\pi/2{\cal B}}\;
\exp\left({\cal A}^2/2{\cal B}^2\right)$ and 
$\aleph_1 \rightarrow -{\cal A}\aleph_0/{\cal B}$.
Taking the limits ${\cal A} \rightarrow 1$, ${\cal B}\rightarrow 0$  
and ${\cal A} \rightarrow 0$, ${\cal B}\rightarrow 1$ in (\ref{pslinPW}) 
one simply finds
exactly the standard Poisson (Eq.\ (\ref{pois})) and 
Wigner (Eq.\ (\ref{wig})) distributions. In this way the constants
${\cal A}$ and ${\cal B}$ are measures of the probability to have 
Poisson and Wigner distributions (Eq.\ (\ref{denlin})). 

\section{NUMERICAL RESULTS}

\begin{figure}
\begin{center}
\includegraphics[width=0.45\textwidth,clip]{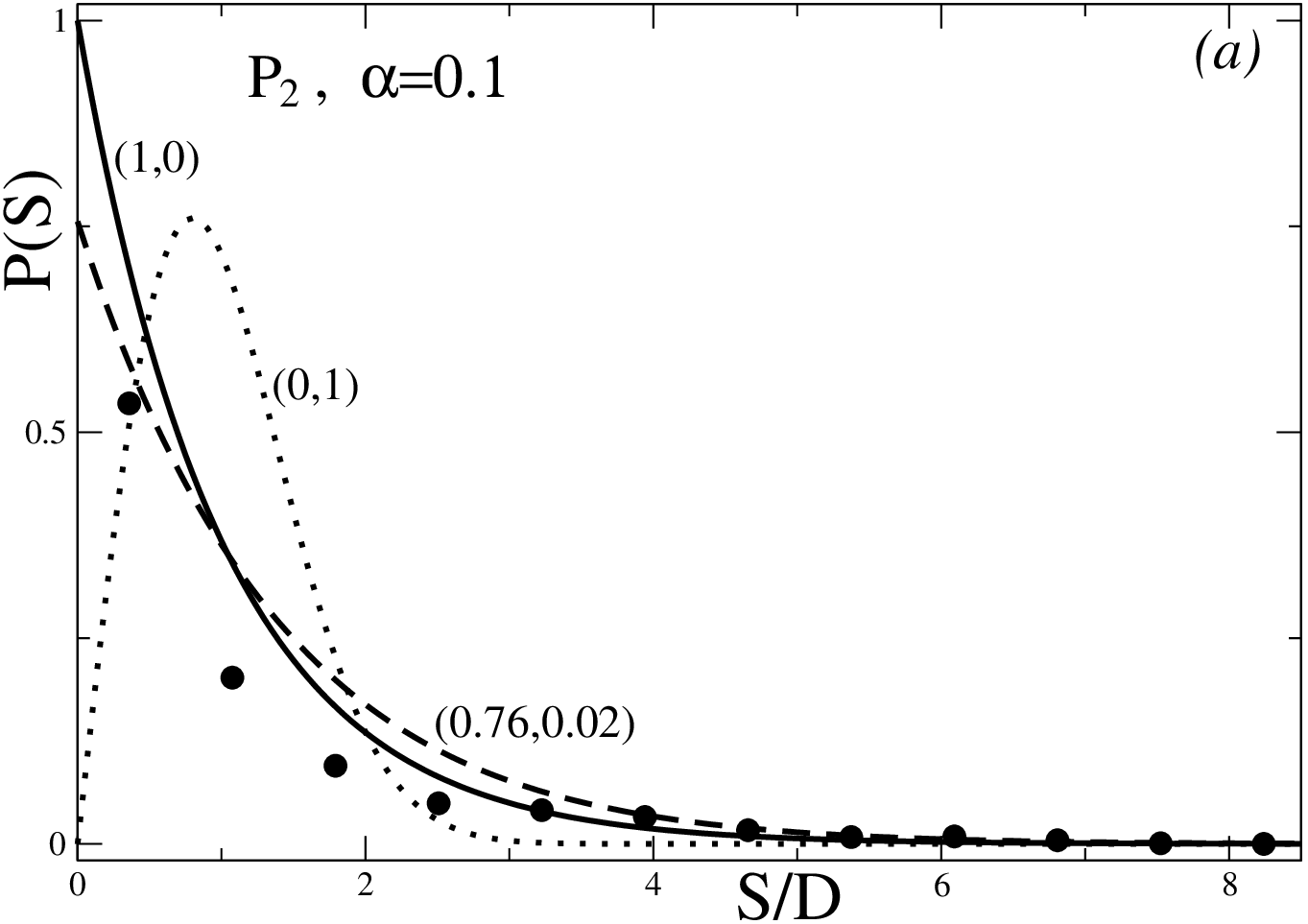}
\hspace{0.2cm}
\includegraphics[width=0.45\textwidth,clip]{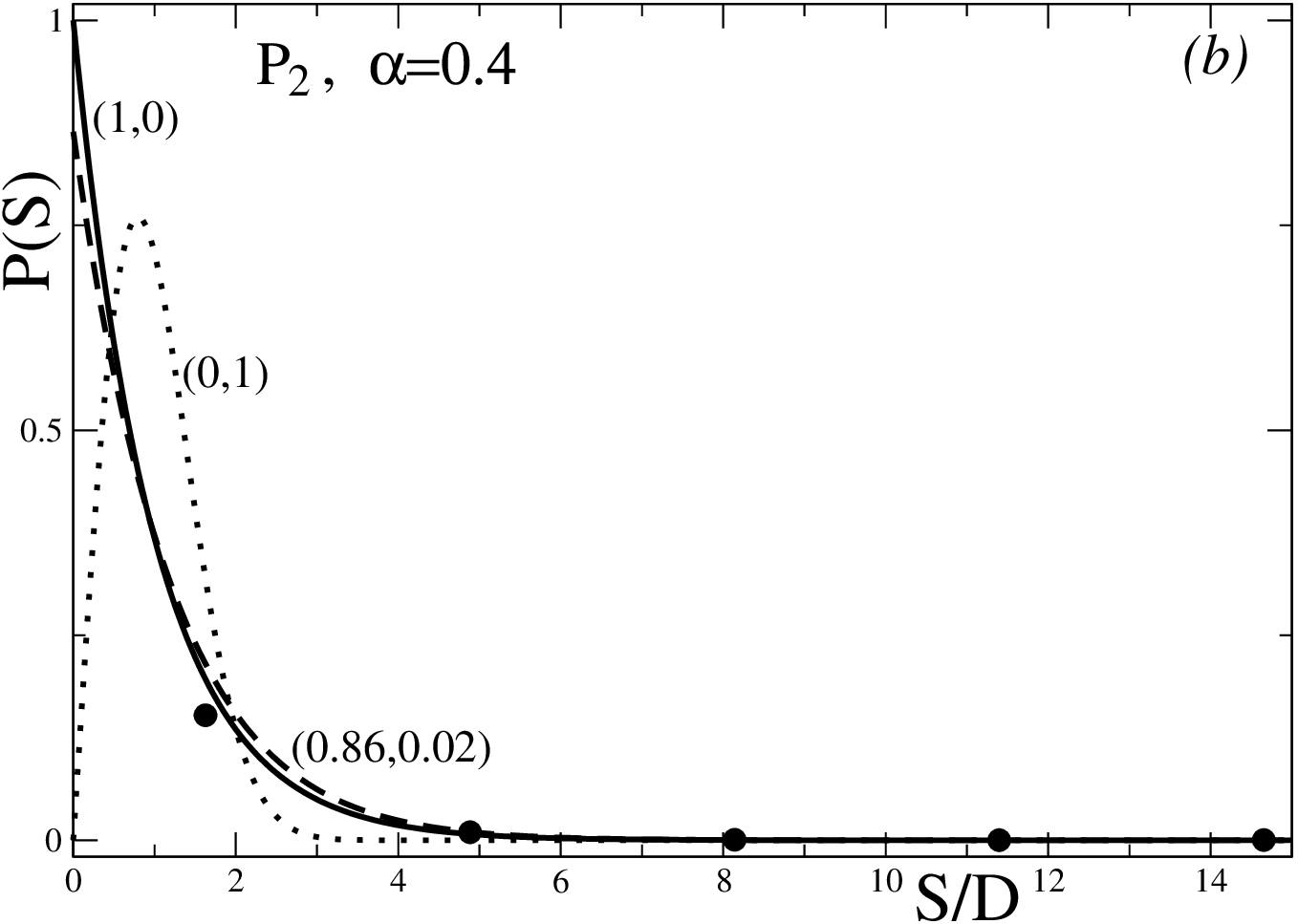}

\includegraphics[width=0.45\textwidth,clip]{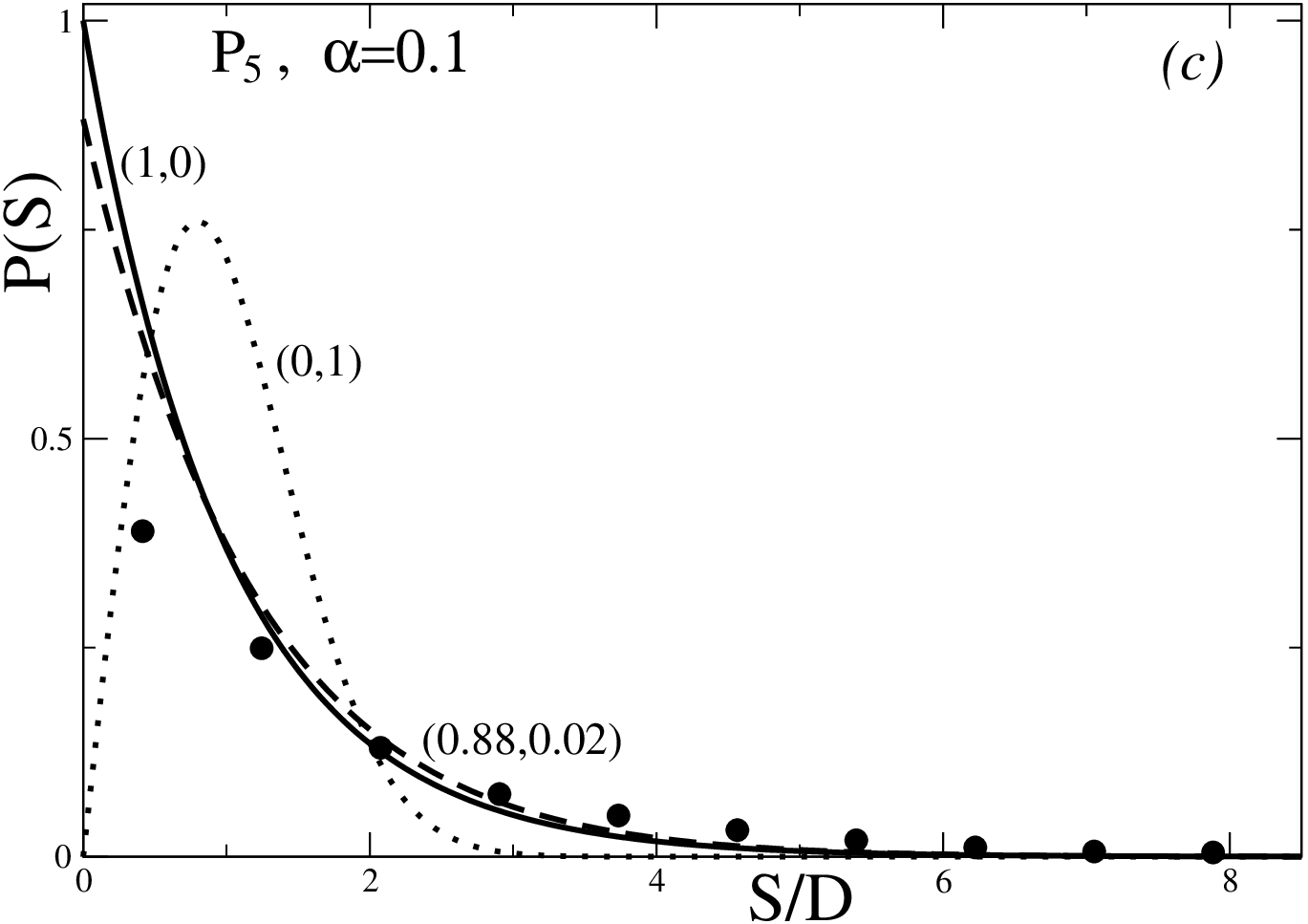}
\hspace{0.3cm}
\includegraphics[width=0.45\textwidth,clip]{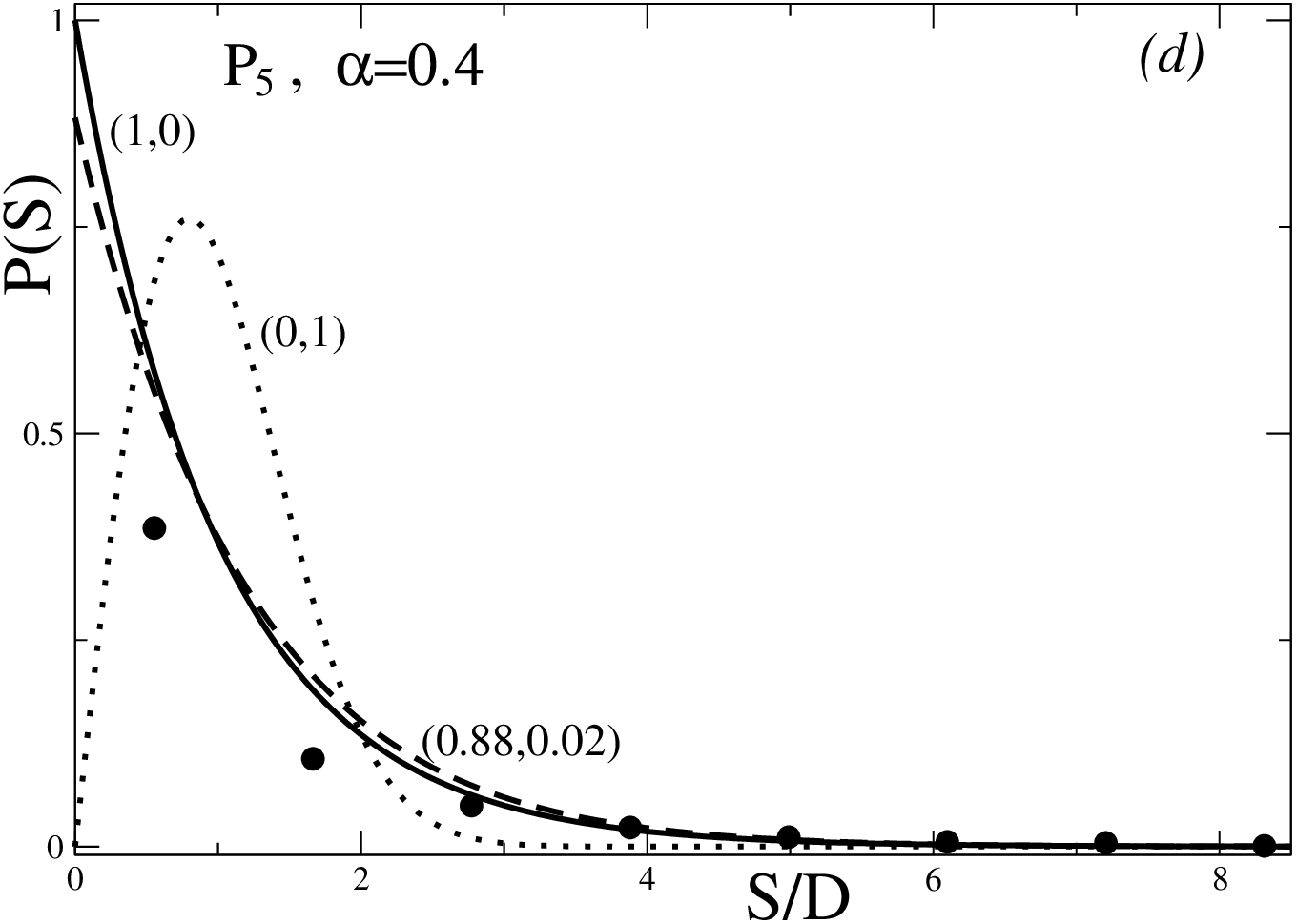}
\end{center}
\caption{ 
The distributions of spacing of the neighboring
 levels $P(S)$ represented by heavy dots \cite{brody} 
vs the energies $S$
for the same spectra as in Figs\ \ref{fig1} and \ref{fig3}. 
Solid curve is a standard Poisson distribution (Eq.\ (\ref{pois})) 
and a dotted one
is a standard Wigner distribution (\ref{wig}). Numbers in brackets
$({\cal A},{\cal B})$ show 
${\cal A}$ and ${\cal B}$ of Eq.\ (\ref{pslinPW}).
Dashed curve corresponds to a linear approximation to the level density 
(Eq.\ (\ref{denlin})); other notations are the same as in 
Figs\ \ref{fig3} and \ref{fig4}.}
\label{fig5}
\end{figure}

\begin{figure}
\begin{center}
\includegraphics[width=0.45\textwidth,clip]{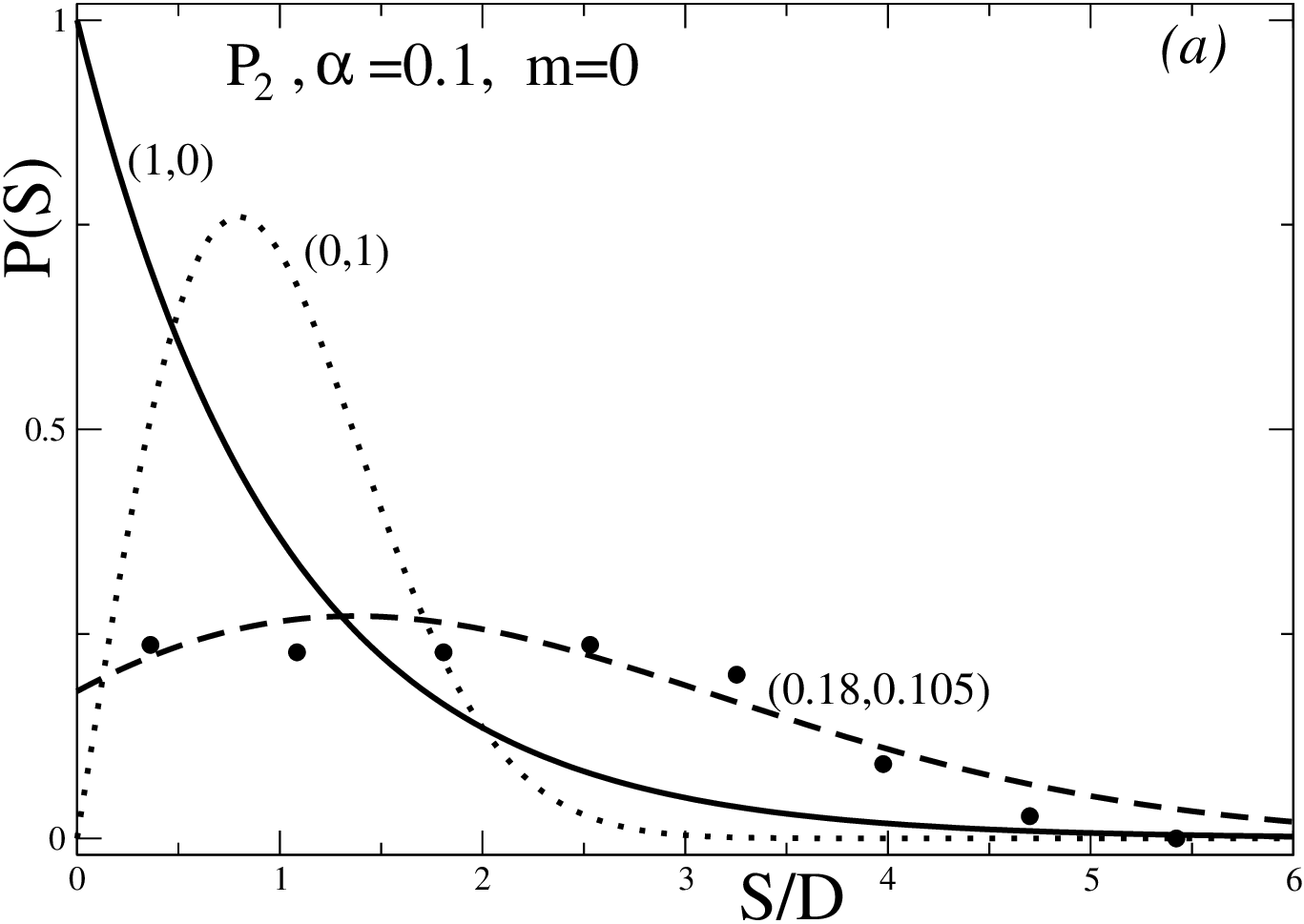}
\hspace{0.3cm}
\includegraphics[width=0.45\textwidth,clip]{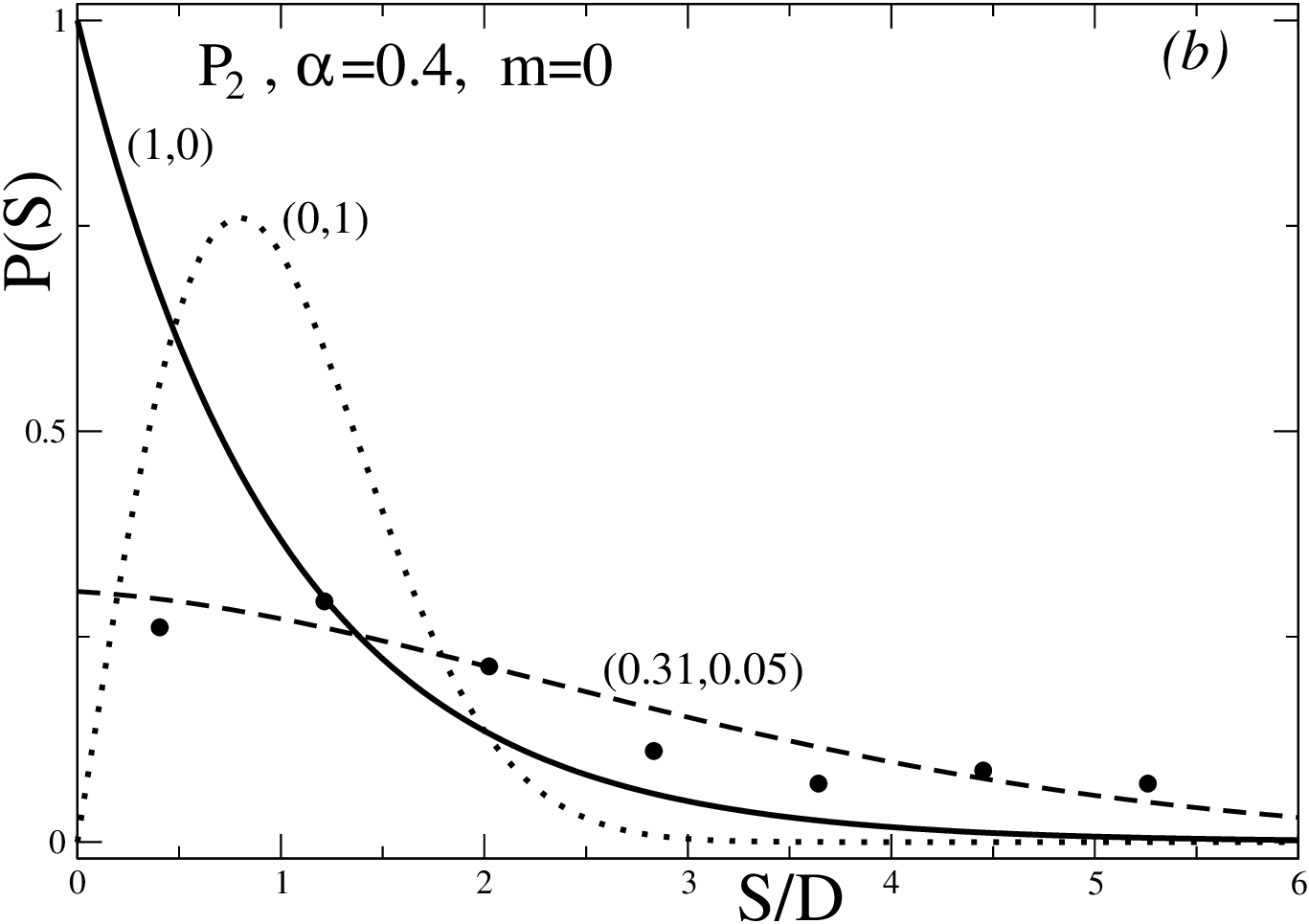}

\includegraphics[width=0.45\textwidth,clip]{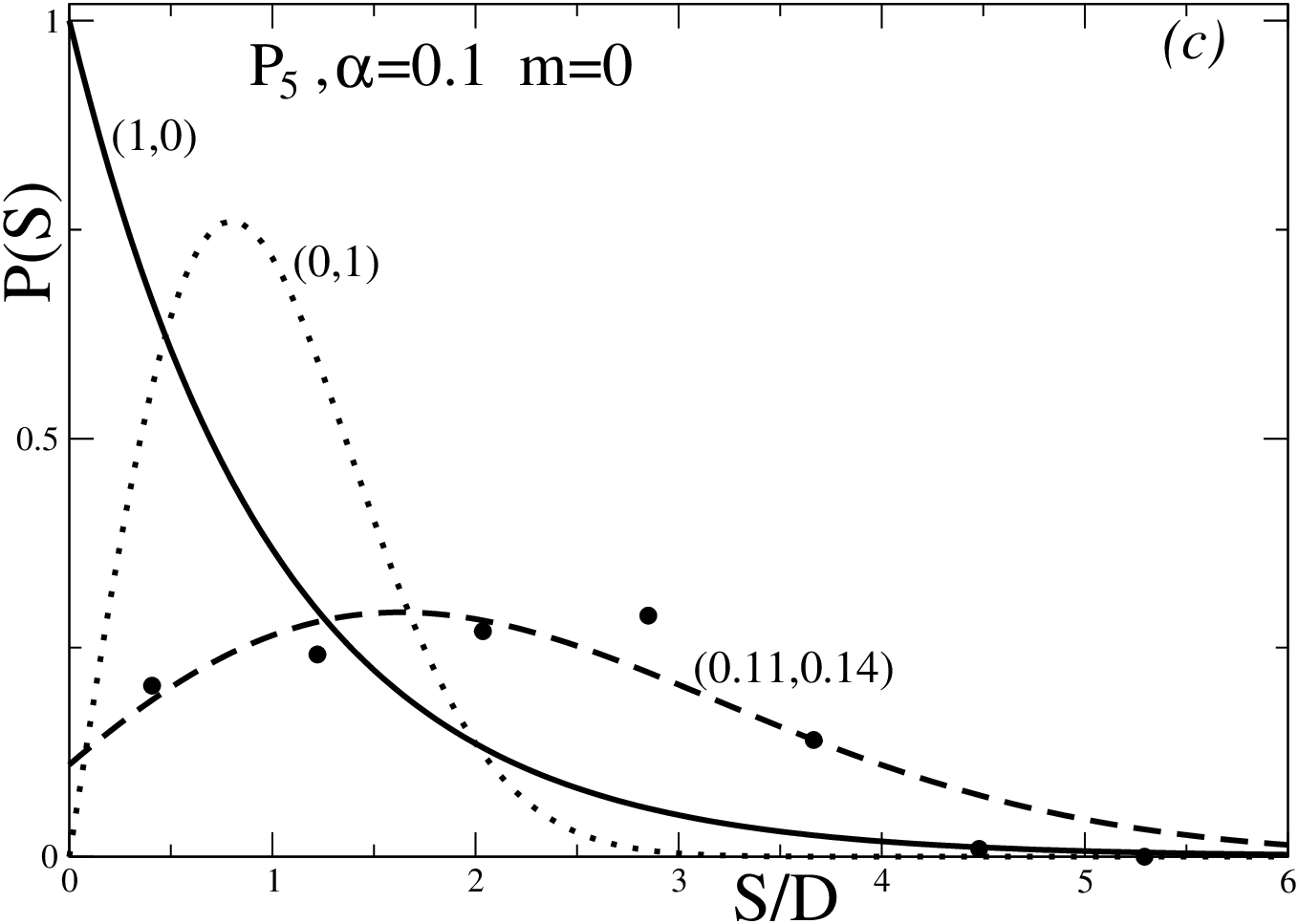}
\hspace{0.3cm}
\includegraphics[width=0.45\textwidth,clip]{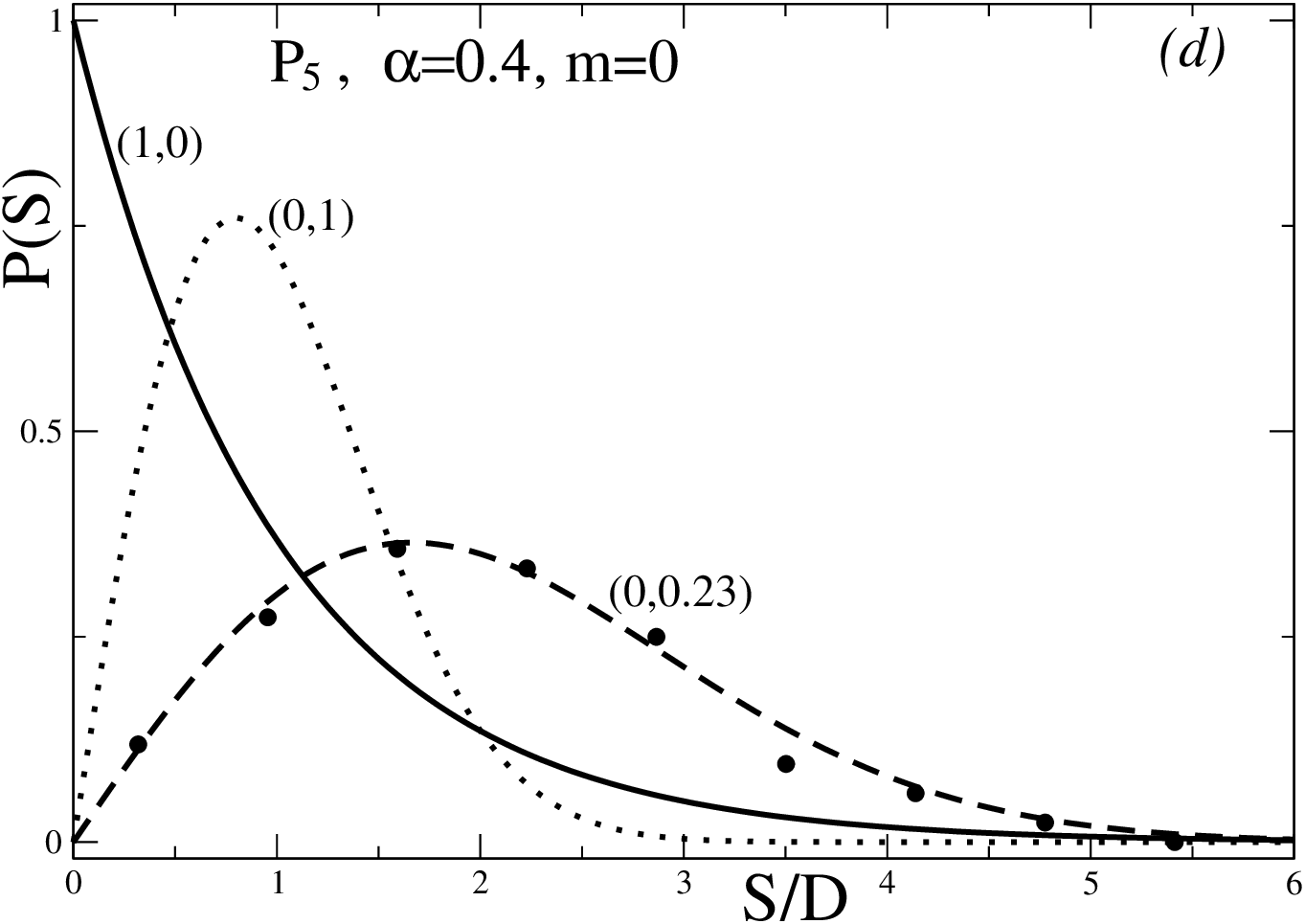}
\end{center}
\caption{ 
The same as in Fig.\ \ref{fig5} but for the spectra of 
Figs\ \ref{fig2}, \ref{fig4}.}
\label{fig6}
\end{figure}

Figs\ \ref{fig5} and \ref{fig6} show the corresponding NNSD 
$P(S)$ (Eq.\ (\ref{genwig})). 
Again, in accordance with spectra (see Figs\ \ref{fig1}, \ref{fig2}) and 
level-density calculations
in Figs\ \ref{fig3} and \ref{fig4}, the dramatic changes are observed between 
Fig.\ \ref{fig6} for the NNSD $P(S)$ with the   
$m=0$ and Fig.\ \ref{fig5}
for those of the full spectra ones. 
Results presented by heavy dots in Fig.\ \ref{fig5} look more close to the 
Poisson distribution and those in Fig.\ \ref{fig6} are more close to the 
Wigner one.

There are
a large difference in numbers ${\cal A}$ and ${\cal B}$ which measure 
the closeness
of the distributions $P(S)$ for the neighboring levels spacing to
the standard 
ones, Poisson (1,0) and Wigner (0,1). However, in Fig.\ \ref{fig6} 
all distributions
are more close to the Wigner in shape having a maximum between zero and 
large compared to $D$  value $S_{max}$ with respect to $D$ [
$S_{max}={\cal C}D$, see immediately 
after Eq.\ (\ref{pslin})] 
than monotonous 
exponential-like decrease similar 
to the Poisson
behavior in Fig.\ \ref{fig5}. Notice that we have more pronounced Wigner-like 
distribution
with increasing multipolarity $n$ and deformation $\alpha$ in Fig.\ \ref{fig6}, 
especially remarkable
at $P_5 $ surface distortions and large enough deformation $\alpha=0.4$, 
see last plot $(d)$ in 
Fig.\ \ref{fig6}. Including all the angular momentum projections $m$ for all 
desired multipolarities and
deformations one has clearly Poisson-like behavior though they differ 
essentially in numbers 
${\cal A},{\cal B}$ from the standard ones (1,0), see Fig.\ \ref{fig6}.

\begin{figure}
\begin{center}
\includegraphics[width=0.9\textwidth,clip]{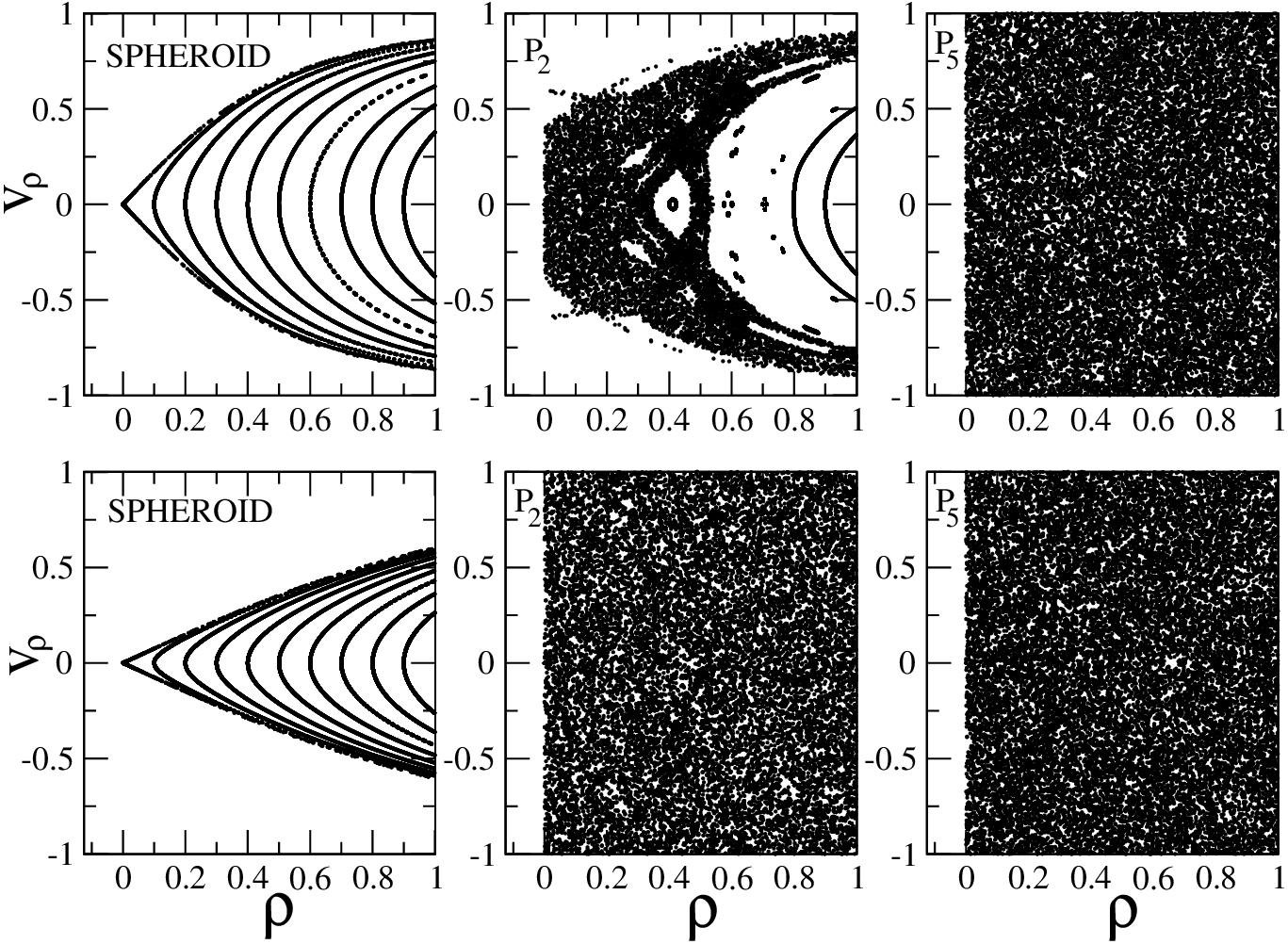}
\end{center}
\caption{
Poincare sections $v_\rho$ vs $\rho$ for 
spheroid, $P_2$ and $P_5$ shapes  
at the small deformation 
$\alpha=0.1$ (upper row) and large deformation 
$\alpha=0.4$ (lower row) for the 
projections of the angular momentum on the symmetry axis $m=0$.}
\label{fig7}
\end{figure}

The reason for this can be understood looking at the 
Poincare sections shown in Fig.\ \ref{fig7} \cite{arvieu,blshsw,kiev2010}. 
The upper row is related to a  small deformation 
and lower row corresponds to a large 
deformation. The projection of the angular momentum is
$m=0$ in all pictures of Fig.\ \ref{fig7}. Difference is remarkable for the 
integrable spheroidal cavity
and other non-integrable (in the plane of the symmetry axis) shapes. As seen
from comparison of upper and lower plot lines,
with increasing deformation $\alpha$ and multipolarity $n$ we find more 
chaotic behavior and we should expect
therefore the NNSD closer to the Wigner distribution (\ref{wig}). This is
in agreement with the NNSD calculations for the fixed $m=0$, see 
Fig.\ \ref{fig6}.  
Notice that similar properties 
of the NNSD for other potentials and 
constraints
were discussed in Refs\ \cite{heiss1,heiss3,nazm1,nazm2}.

The difference between the NNSD calculations 
with the realistic level densities by the Strutinsky 
shell-correction method (see Eq.\ (\ref{totlevden})) for  the considered 
WS potential and those with 
their idealistic linear behavior (Eq.\ (\ref{denlin})) can be
studied in terms of the general formula $P(S)$ 
(Eq.\ (\ref{genwig})). In particular, the shell effects related
to the inhomogeneity of the s.p. levels near the Fermi surface for all 
desired
multipolarities and deformations are found to be significant, 
also in relation to the fixed  
quantum number $m$. 

\begin{figure}
\begin{center}
\includegraphics[width=0.45\textwidth,clip]{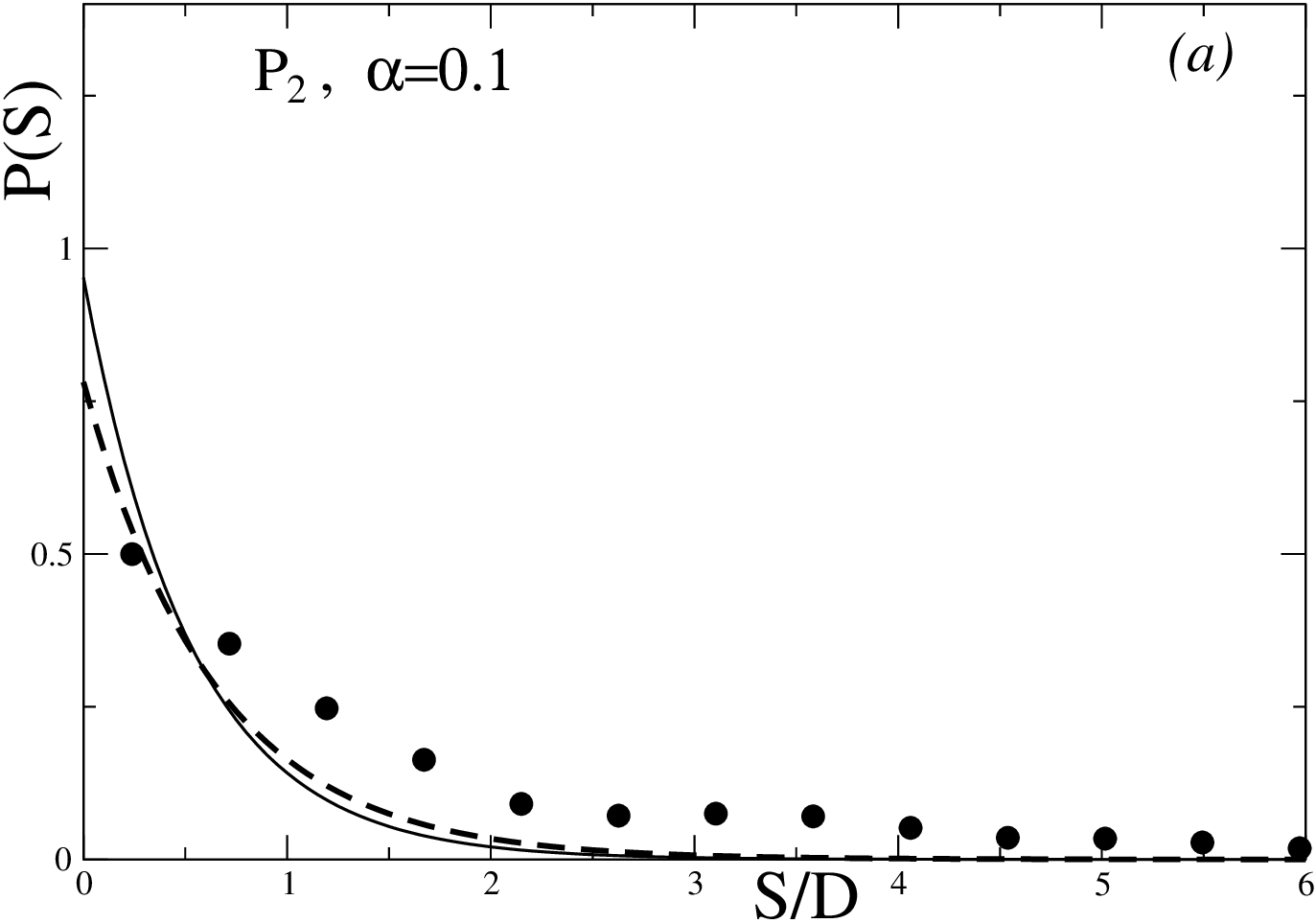}
\hspace{0.3cm}
\includegraphics[width=0.45\textwidth,clip]{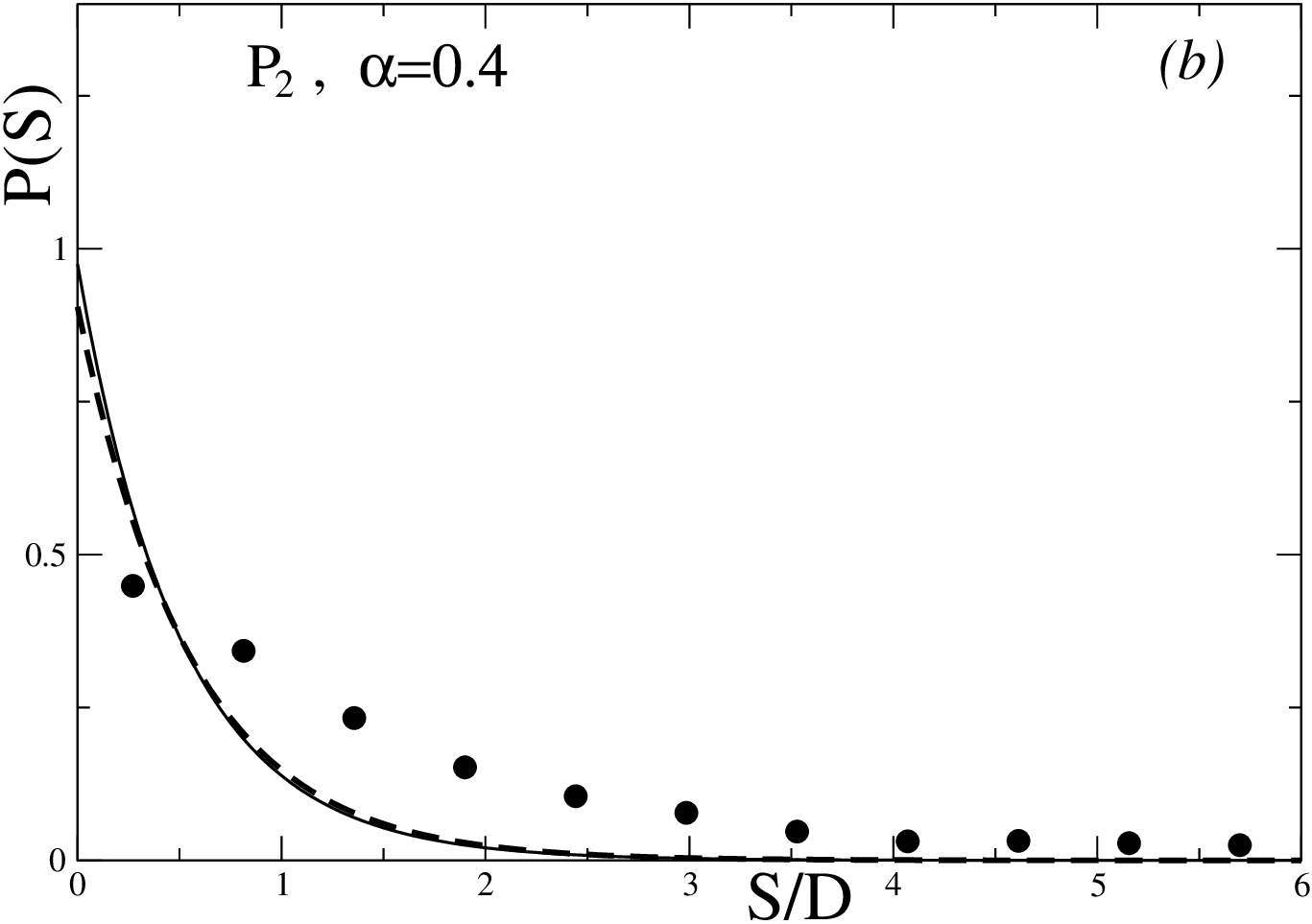}

\includegraphics[width=0.45\textwidth,clip]{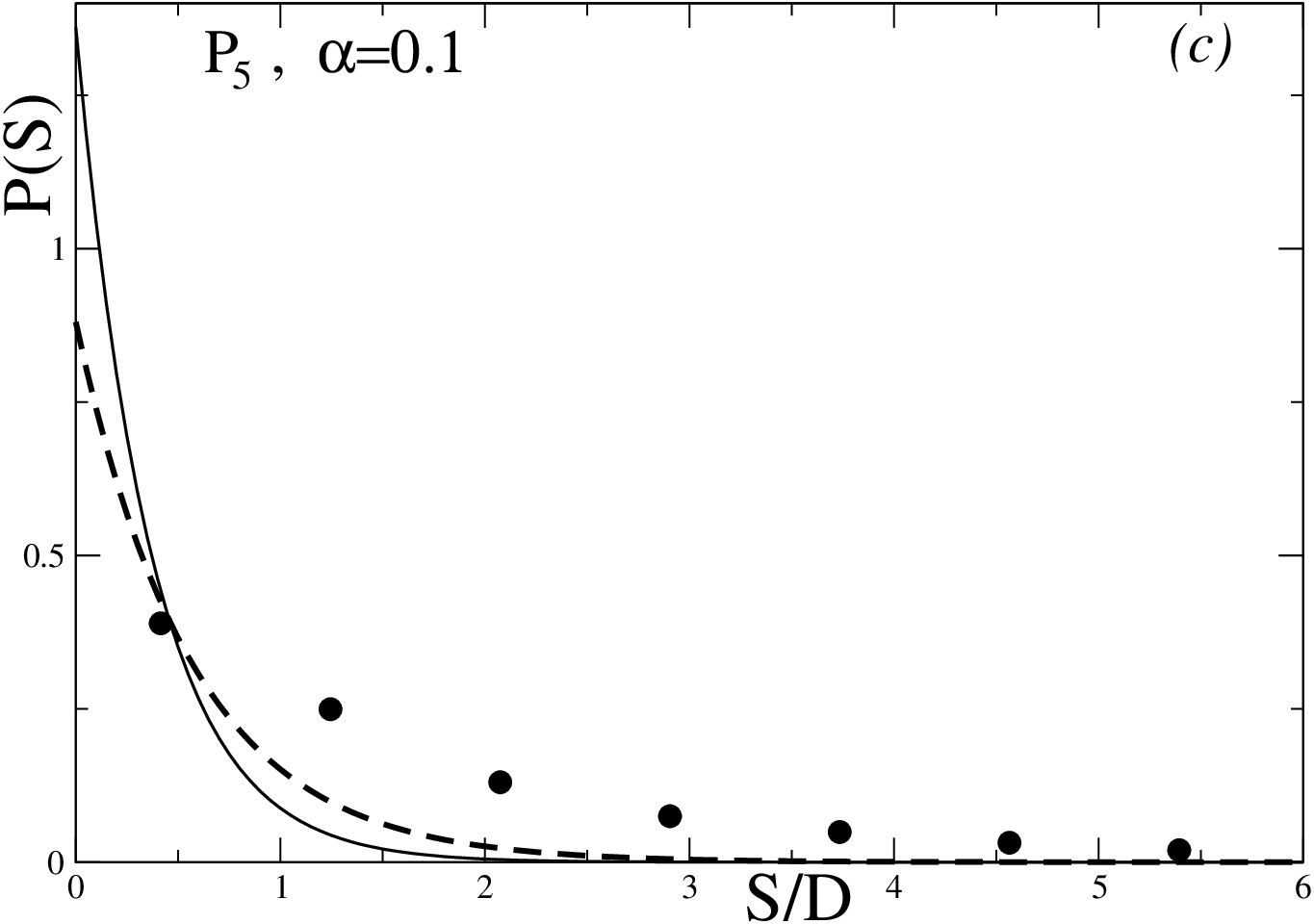}
\hspace{0.3cm}
\includegraphics[width=0.45\textwidth,clip]{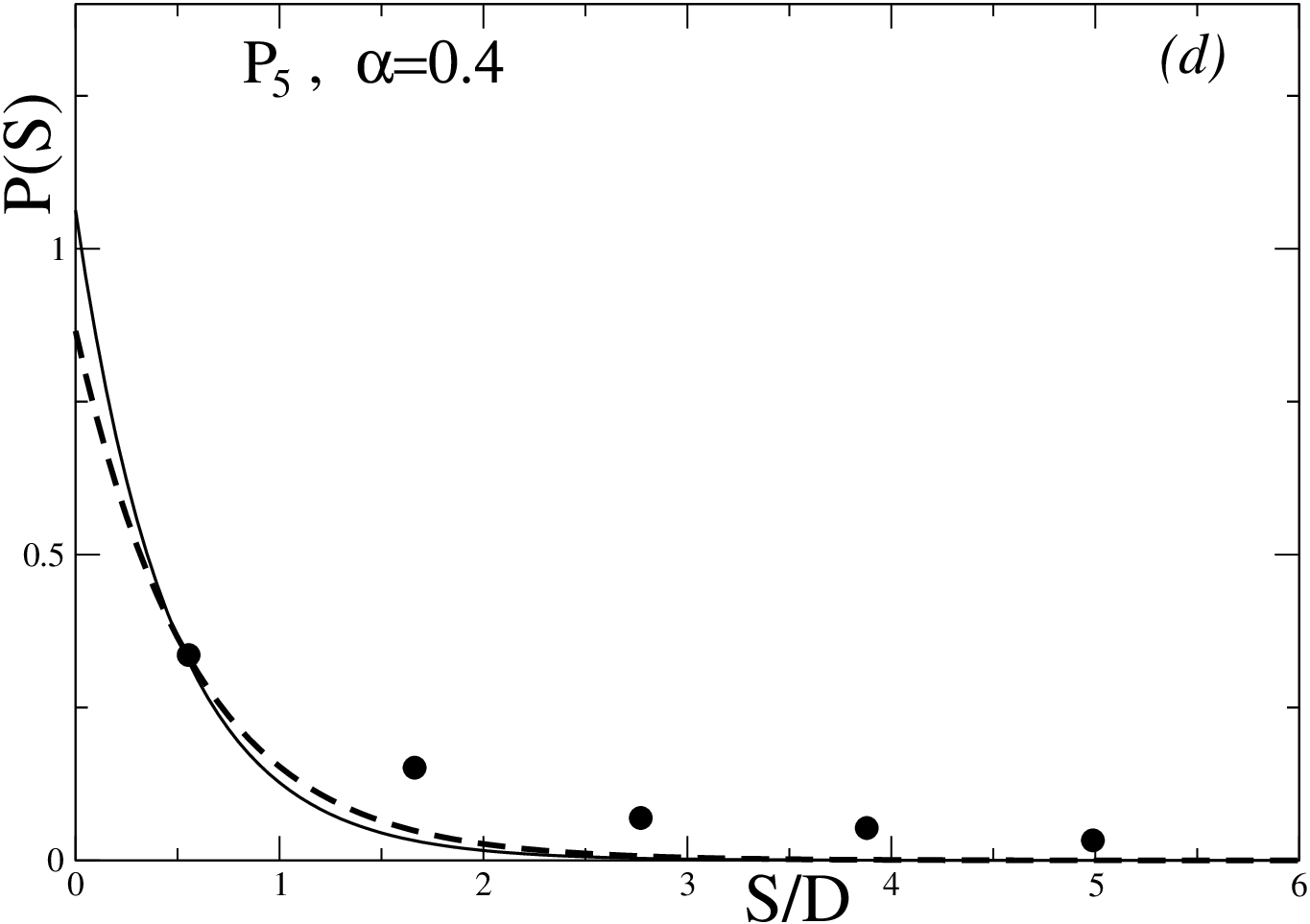}
\end{center}
\caption{
The  general 
distribution $P(S)$ (Eq.\ (\ref{genwig}))  
vs the energies $S$
for the same spectra as in Figs\ \ref{fig1}, \ref{fig3}, \ref{fig5}; 
dashed are the 
distributions $P(S)$ related to the
Strutinsky smooth density $\tilde{g}(S)$ and 
solid is the total level density $g_\Gamma(S)$ (Eq.\ (\ref{totlevden})); 
dots are the same
as in Figs\ \ref{fig5} and \ref{fig6}.}  
\label{fig8}
\end{figure}

\begin{figure}
\begin{center}
\includegraphics[width=0.45\textwidth,clip]{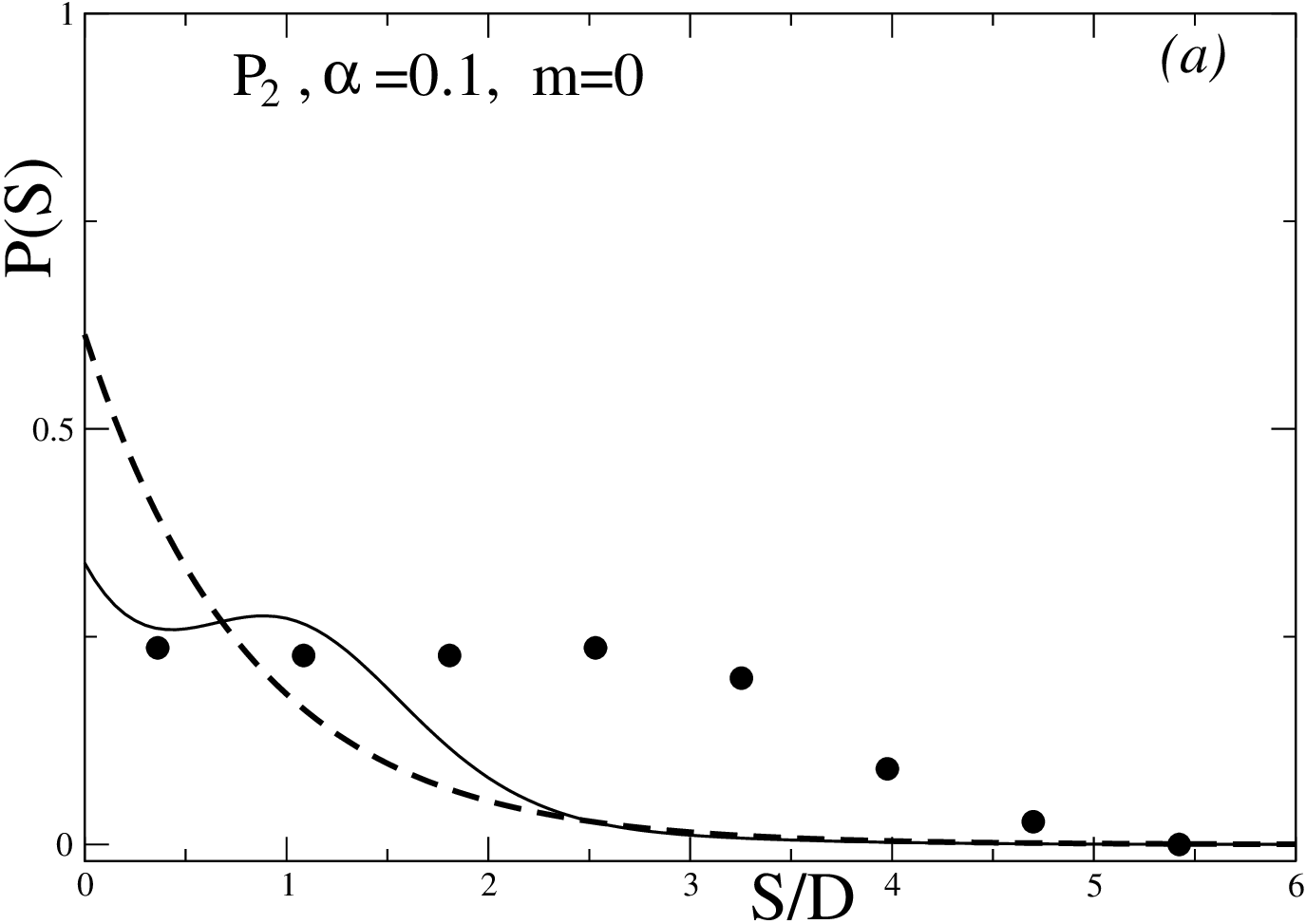}
\hspace{0.3cm}
\includegraphics[width=0.45\textwidth,clip]{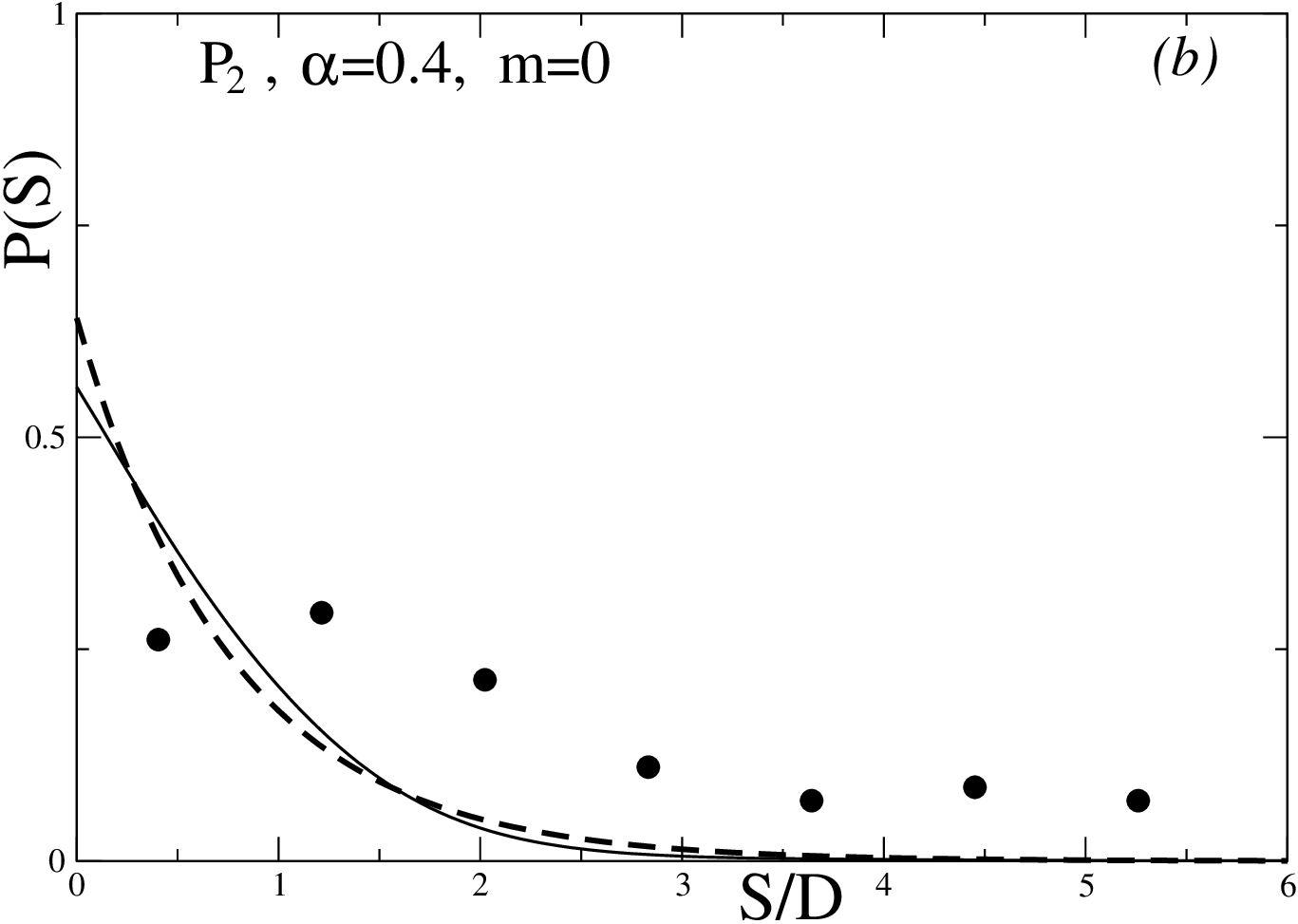}

\includegraphics[width=0.45\textwidth,clip]{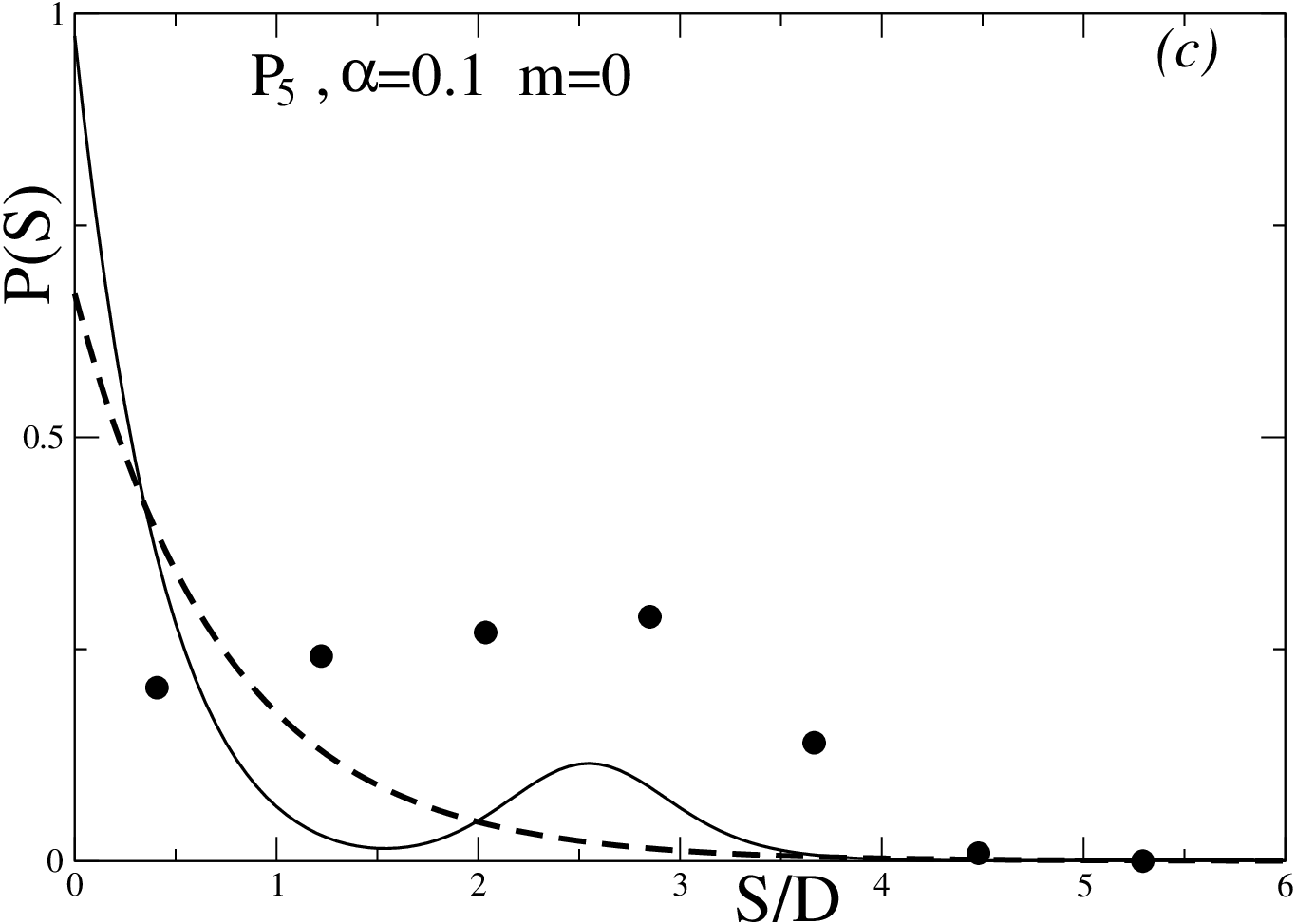}
\hspace{0.3cm}
\includegraphics[width=0.45\textwidth,clip]{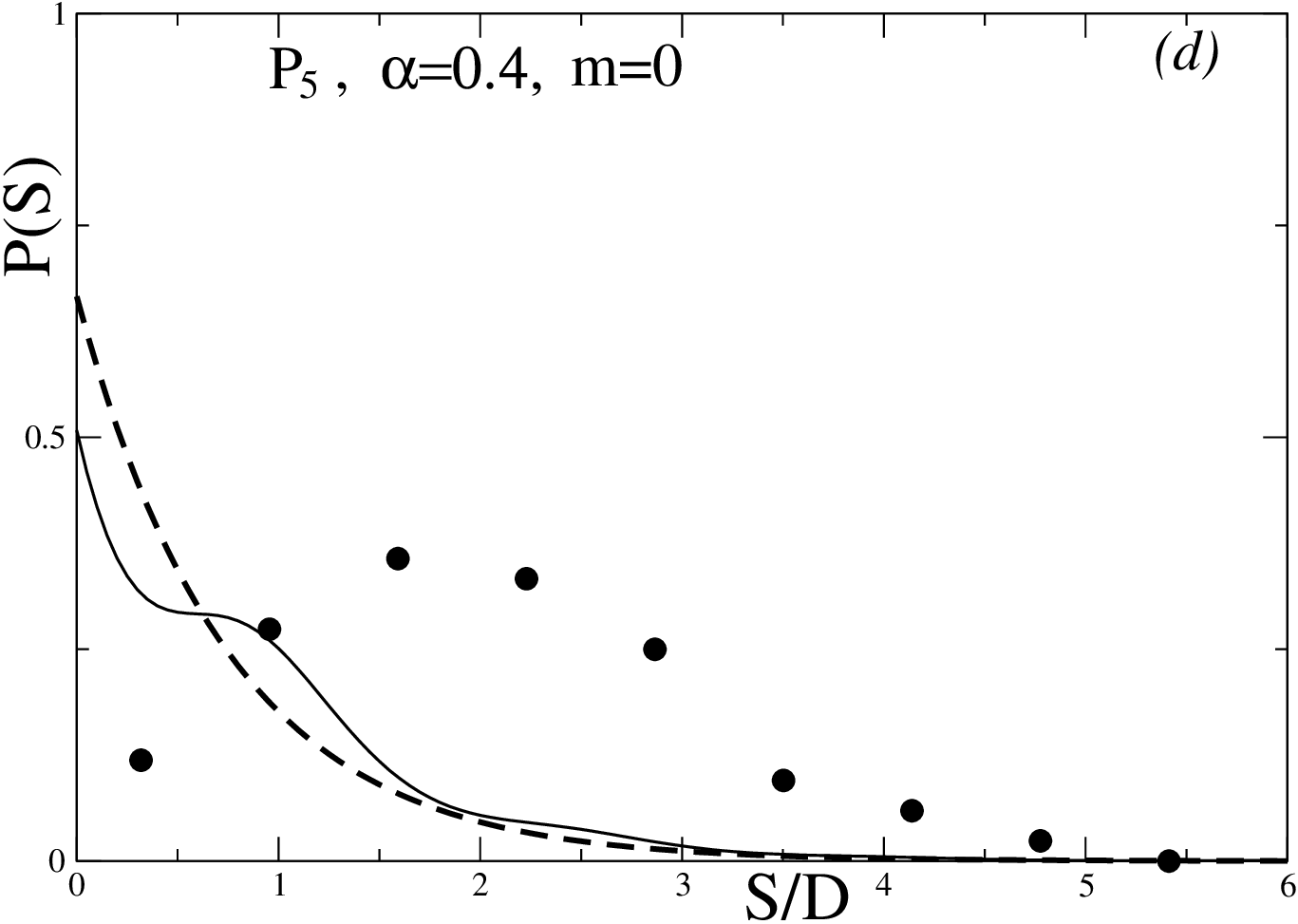}
\end{center}
\caption{ 
The same as in Fig.\ \ref{fig8} but for spectra of 
Figs\ \ref{fig2}, \ref{fig4} and \ref{fig6}.}
\label{fig9}
\end{figure}

Figs\ \ref{fig8} and \ref{fig9} show 
the results of these calculations corresponding 
to Figs\ \ref{fig5} and \ref{fig6}.
Notice that in the case of the full spectra, see Fig.\ \ref{fig8}, 
one has Poisson-like
distributions corresponding to the smooth density (dashed) with a 
similar behavior as NNSD shown by heavy dots, 
in contrast to
Fig.\ \ref{fig9} where we find rather big differences between these curves.
The shell effects are measured by the differences between the solid 
curve related to
 the total level density with the shell components and the 
dashed one for the smooth
level density of Figs\ \ref{fig3} (all $m$) and 
\ref{fig4} (with $m=0$), see correspondingly 
Figs\ \ref{fig8} and \ref{fig9}.
With increasing deformations one has slightly decreasing the shell effects, 
in contrast to
the multipolarity dependence for which there is almost no change of the 
shell effects
at the same deformations.

\section{CONCLUSIONS}
\label{concl}

We studied the statistics of the neighboring s.p. levels in the WS potential 
for several 
typical multipolarities and deformations of the surface shapes and 
deformations, as 
compared with the standard Poisson and Wigner distributions $P(S)$. 
For the sake of comparison,
we derived analytically the combine asymptotic 
Poisson-Wigner distribution $P(S)$ related to the general linear
dependence of the corresponding s.p. level density. We found the significant
differences between distributions for a fixed value of the 
angular momentum
projection $m$ of the particle and those accounting all 
possible values of $m$. 
 For the case of the fixed $m=0$ we obtained distributions
$P(S)$ more close to the Wigner shape with the maximum between $S=0$ 
and a maximal large 
value of $S$, the more pronounced the larger multipolarity and deformation
of the potential surface. 
We found also that the full spectra distributions
$P(S)$ look Poisson-like in a sense that they have maximum at $S=0$ 
and almost exponential
decrease as a function of the energy near the Fermi surface. 
Our results clarify the widely extended opinion 
of the relation of the distributions 
(Poisson or Wigner) to the integrability of the problem 
(the integrable or chaotic one). All
considered potentials are axially-symmetric but they are the same
{\it non-integrable} ones in the plane of the
symmetry axis. However, the degree of the symmetry 
(classical degeneracy \cite{myab,strutmag,smod} 
${\cal K}$, i.e. the number of 
the single-valued integrals of motion besides
of the energy) for the case of the full spectra, Figs\ \ref{fig1}, 
\ref{fig3}, \ref{fig5}, 
( ${\cal K}=1$, a mixed system) is higher than 
for the fixed angular momentum $m$ (${\cal K}=0$ like
for the completely chaotic system).  
Notice that integrability is not only one criterium of
chaoticity.
The measure of the differences of the distributions $P(S)$ 
between Poisson and Wigner standard ones depends also
on the properties of the energy dependence of the 
level density (from constant to proportional-to-energy dependence). 
From comparison
between the general 
distribution $P(S)$ related to the smooth level density obtained by the 
Strutinsky shell correction method and the statistics of the neighboring 
s.p. levels one finds  that all of them are more 
close to the Poisson-like behavior. This shows that the energy dependence
of the smooth level density differs much from the linear functions. 
We obtained also
large shell effects in the distributions $P(S)$ in nice 
agreement with those in the
key quantity in this analysis, - level density dependencies on 
the energy near the Fermi surface.

As to perspectives,
it might be necessary to use the
combined microscopic-macroscopic approaches \cite{myersswiat,strut}
to clear up the results more systematically and analytically. 
Our quantum results can be interesting for 
understanding the one-body dissipation at  
slow and faster collective dynamics with different shapes
like the ones met in 
nuclear fission and heavy-ion collisions.

\section{Acknowledgements}

We thank S. Aberg, V.A. Plujko and  S.V. Radionov
 for valuable discussions.

\appendix \setcounter{equation}{0}
\renewcommand{\theequation}{A\arabic{equation}}

\begin{center}
\textbf{Appendix A: A derivation of the NNSD}
\label{appA}
\end{center}
 
We introduce first the level density, $g(E)$, as the number of the 
levels  ${\mbox d} N$ in the energy interval [$E,E+{\mbox d} E$]
divided by the energy interval, $g(E)={\mbox d} N/{\mbox d} E$. 
With the help of this
quantity one can derive the 
NNSD $P(S)$ as the probability density versus
the spacing $S$ between the nearest neighboring levels. Specifying $P(S)$
to the problem with the known s.p. spectra of the Hamiltonian, one can 
split the energy interval $\Delta E$ under the investigation into many small
(equivalent for simplicity) parts $\Delta S \ll \Delta E$. 
Each of $\Delta S$ nevertheless contain 
many energy levels, $\Delta S \gg D$. 
Then, we find the number of the levels  
which occur inside of the 
small interval $\Delta S$.
Normalizing these numbers by the total number of the levels
inside the total energy interval $\Delta E$ one obtains the distribution
which we shall call as the probability 
density $P(S)$. Notice that the result of this calculation depends on 
the energy length of the selected $\Delta S$. In our calculations, 
we select $\Delta S$ by the condition of a sufficient smoothness of the
distribution $P(S)$. Such procedure is often used for the statistical treatment
of the experimentally obtained spectrum with the fixed quantum numbers 
like the angular momentum, parity and so on \cite{brody}. 

Following mainly Ref.\ \cite{aberg}, let us calculate first the 
intermediate quantity 
$f(S)$ as the probability that there is no energy level 
in the energy interval [$E,E+S$]. According to the general 
definition of the level density
mentioned above,  $g(S) {\mbox d} S$ can be considered as the probability 
that there is one energy level in 
$[E+S, E+S+{\mbox d} S]$. Then, 
\begin{equation}
f(S+{\mbox d} S)=f(S)\left(1-g(S) {\mbox d} S\right)\;,
\label{f}
\end{equation}
which leads to the differential equation for $f(S)$,
\begin{equation}
{\mbox d} f= -g(S) {\mbox d} S f(S).
\label{diff}
\end{equation}
Solving this equation one gets
\begin{equation}
f(S)=C \exp\left(-\int_0^S g(x) {\mbox d} x\right)\;.
\label{fsol}
\end{equation}
Let $P(S) \mbox{d} S$ denote the probability that the next energy level is in 
$[E+S, E+S+\mbox{d} S]$,
\begin{equation}
P(S) {\mbox d} S = f(S) g(S) {\mbox d} S\;.
\label{defps}
\end{equation}
 Then, substituting Eq.\ (\ref{fsol}) into Eq.\ (\ref{defps}) one finally 
arrives at the general distribution:
\begin{equation}
P(S)=C g(S)  \exp\left(-\int_0^S g(S') {\mbox d} S'\right).
\label{psol}
\end{equation}
The boundary conditions in solving the differential equation (\ref{diff})
accounts for the meaning of the NNSD
$P(S)$ and its argument as the spacing between the nearest neighbor levels
as shown in the integration limit in Eq.\ (\ref{psol}). The constant $C$
is determined from the normalization conditions (Eq.\ (\ref{norm})).

%
%
%
%
%
%
%
%
%
\end{document}